\documentclass[lettersize,journal]{IEEEtran}
\usepackage{amsmath,amsfonts}
\usepackage{array}
\usepackage[caption=false,font=normalsize,labelfont=sf,textfont=sf]{subfig}
\usepackage{textcomp}
\usepackage{stfloats}
\usepackage{url}
\usepackage{verbatim}
\usepackage{graphicx}
\usepackage{svg}
\svgpath{{images/}}
\usepackage{wrapfig}
\usepackage{floatpag}
\graphicspath{ {./images/} }
\usepackage{cite}
\def\BibTeX{{\rm B\kern-.05em{\sc i\kern-.025em b}\kern-.08em
    T\kern-.1667em\lower.7ex\hbox{E}\kern-.125emX}}
\usepackage{balance}
\usepackage{algorithm}
\usepackage{algpseudocode}
\usepackage{xcolor}
\usepackage{makecell}
\usepackage{soul}

\begin{document}

\title{Safely Mitigating Grid Imbalance: A Hybrid Distributional Wide-Margin Recourse Approach}

\author{Shoaib Hussain, \emph{Student Member, IEEE}, Mostafa Farrokhabadi, \emph{Senior Member, IEEE}, Hamidreza Zareipour, \emph{Fellow, IEEE}
\thanks{This paper was produced by the IEEE Publication Technology Group. They are in Piscataway, NJ.}
\thanks{Manuscript received April 19, 2021; revised August 16, 2021.}}

\markboth{Journal of \LaTeX\ Class Files,~Vol.~14, No.~8, August~2021}%
{Shell \MakeLowercase{\textit{et al.}}: A Sample Article Using IEEEtran.cls for IEEE Journals}


\maketitle

\begin{abstract}
This paper proposes the Multi-Agent Distributional Wide-Margin Policy Gradient (MADWMPG) framework: a dual-policy hybrid imitation and safe reinforcement learning (IL-SRL) method designed for safely mitigating grid imbalance in multiphase active distribution networks with distributed soft open points (SOPs). The framework comprises a wide-margin constraint barrier recourse (WCBR) module that seeks the maximal separation of constraint-compliant and unsafe actions and learns to proactively suppress exploratory violations. The framework addresses operational volatility using a novel adaptation of noncrossing quantile regression (NQR) distributional critics that track the complete distribution of policy returns. The set of behavior and recourse policies integrates an instantaneous physics-based and differentiable action projection mechanism that generates SOP constraint-compliant actions without relying on active learning. Policy safety is made responsive using an enhancement of conventional experience replay and the safety performance of the framework is supported by probabilistic bounds derived in the paper. The proposed framework was trained and tested on a modified multiphase IEEE 123-bus test feeder. Results demonstrate its superiority to conventional reinforcement learning and SRL methods and show that it is highly competitive against the nonlinear AC optimal power flow. 
\end{abstract}



\section{Introduction}
\IEEEPARstart{T}{he} rapid proliferation of intermittent distributed energy resources (DERs), coupled with inherently asymmetric phase loadings and stochastic demand profiles \cite{zanjaniNovelApproachMitigating2025}, has exacerbated operational unbalance in modern distribution networks. As residential rooftop photovoltaic (PV) arrays, localized battery storage units, and plug-in electric vehicle (EV) charging infrastructures \cite{louNewPhasechangingSoft2020a} are unevenly integrated across low-voltage secondary feeders, the network exhibits severe spatial and temporal load asymmetries across phases. This three-phase voltage and current imbalance induces negative and zero-sequence components \cite{salehCoordinatedUnbalanceCompensation2024} that drive elevated neutral conductor currents, excessive power losses along line conductors, localized distribution transformer core saturation, and counter-torque overheating in electrical motors \cite{chuanjiangExperimentalStudyControlled2017}. Concurrently, persistent phase unbalances may breach statutory voltage unbalance factor (VUF) thresholds \cite{rahmanNewApproachVoltage2018}, erode network renewable hosting capacity, and cause premature insulation breakdown in grid assets \cite{kimExaminationEffectReactive2020}. Conventional grid mitigation schemes such as manual phase re-balancing, discrete step voltage regulators, switched capacitor banks (SCBs), and mechanical tie-switches are constrained by operational response delays as well as finite mechanical endurance, rendering them incapable of tracking high-frequency power fluctuations \cite{caoBenefitsAnalysisSoft2016, chethanReviewFACTSDevice2024}. Consequently, network imbalance has evolved from a marginal concern to a mission-critical operational requirement to maintain grid efficiency and system reliability. To address the intensifying power quality concerns, system operators are increasingly exploring distribution-level Flexible AC Transmission Systems (D-FACTS) such as Soft Open Points (SOPs) in pursuit of dynamic mitigation strategies \cite{jiangOverviewSoftOpen2022a}. 

Soft open points are advanced power-electronic devices installed at normally open points (NOPs) between distribution feeders to provide flexible, decoupled, and continuous four-quadrant active and reactive power control \cite{caoOperatingPrincipleSoft2016}. SOPs replace mechanical tie-switches with back-to-back voltage-source converters (B2B VSCs) linked by a shared DC bus \cite{jiangOverviewSoftOpen2022a}. Through this decoupled continuous control, conventional SOPs alleviate feeder congestion, smooth voltage fluctuations \cite{liDdpgBasedAdaptiveVoltage2021}, minimize system losses, and facilitate post-fault service restoration far more effectively than mechanical switches \cite{wangDatadrivenStochasticService2020}. However, conventional symmetrical SOP topologies operate under an identical phase connection sequence, which fundamentally prevents them from transferring active power across disparate phases and addressing the root cause of grid imbalance. To overcome this inter-phase transfer bottleneck, the phase-changing soft open point (PCSOP) enables direct cross-phase power routing \cite{louNewPhasechangingSoft2020}. The decoupled reactive power injection allows PCSOPs to dynamically enforce symmetrical voltage vectors across grid nodes whilst simultaneously modulating the inter-phase bulk powerflow at the substation root bus \cite{louMultiterminalPhasechangingSoft2022}. Nevertheless, capitalizing on the high-speed operational flexibility of PCSOPs requires dispatch frameworks capable of handling the non-convex physics of multi-phase electrical networks. 

Optimal dispatching in active distribution networks (ADNs) has traditionally relied on mathematical programming techniques which struggle to satisfy the convergence and scalability demands of real-time operations \cite{yangOptimalReactivePower2017}. The baseline problem formulation of multi-period AC optimal power flow (ACOPF) involving SOPs is nonlinear and non-convex. Furthermore, the combination of ACOPF with the complex SOP operational dynamics yields a solution search space that is challenging to navigate \cite{gaoDatadrivenMultimodeAdaptive2024}. Incorporating the discrete decision variables of legacy assets alongside SOPs elevates the problem to high-dimensional mixed-integer nonlinear programs (MINLP) that further complicates tractability and solution times \cite{baiDistributedEnergyStorage2018}. To circumvent non-convexity, prior art on SOP applications widely employs convex relaxation techniques to transform the MINLP formulations into linear and mixed-integer linear programming (MILP) solutions \cite{louMultiterminalPhasechangingSoft2022,louNewPhasechangingSoft2020}. However, the physical exactness of solution relaxations cannot be guaranteed in realistic unbalanced distribution networks characterized by asymmetric phase couplings. Additionally, statutory voltage limits impose varying levels of optimization difficulty depending on regional operating codes. Under tight operational constraints, optimization solvers suffer from numerical ill-conditioning and computational delays \cite{louMultiterminalPhasechangingSoft2022}, rendering them unsuitable for continuous online dispatch.                 

The limitations of classical optimization frameworks and their dependency on exact physical models have spurred researchers toward data-driven frameworks. Advances in compute, pattern recognition and pervasive advanced metering infrastructure has led to a marked resurgence in adoption of adaptive sequential decision-making frameworks such as Deep Reinforcement Learning (DRL) \cite{chenReinforcementLearningSelective2022}. DRL formulates the network operation problem as a Markov Decision Process (MDP) and constructs the parameterized functional mappings directly from observed system states to actuator setpoints \cite{glavicDeepReinforcementLearning2019a}. Unlike model-based optimization frameworks, the DRL inference can be executed online in milliseconds. This allows the closed-loop DRL agents to keep up pace with volatile DER ramp rates and demand fluctuations that might overwhelm traditional numerical approaches \cite{zhangDDPGBasedMultiAgentFramework2021a}.

DRL methods have been studied extensively for applications spanning Volt-VAR optimization \cite{leeGraphPolicyNetwork2022a}, energy storage coordination \cite{caoDeepReinforcementLearningBased2020a}, network reconfiguration \cite{yinSequentialReconfigurationUnbalanced2023a}, heterogeneous multi-timescale dispatch \cite{yangTwoTimescaleVoltageControl2020a}, and grid restoration \cite{zhaoLearningSequentialDistribution2022a}. The applications have steadily progressed from simple tabular methods \cite{glavicReinforcementLearningElectric2017a} to advanced multi-agent actor-critic frameworks \cite{suReviewSafeReinforcement2025}. Earlier applications relied on Deep Q-Networks (DQN) to optimize the discrete tap positions of voltage regulators and switched capacitor banks \cite{chenReinforcementLearningSelective2022}. Applications that demand continuous control largely focus on policy gradient and actor-critic algorithms; most notably the Deep Determininistic Policy Gradient (DDPG), Soft Actor Critic (SAC), Twin Delayed DDPG (TD3PG), and Proximal Policy Optimization (PPO) \cite{suReviewSafeReinforcement2025}. A search on DRL-based SOP applications returns only a handful of results. For example, a two-time scale method is proposed in \cite{xiongReactivePowerOptimization2024a} where the hourly dispatches for on-load tap-changing voltage regulators and SCBs are first obtained using an MILP method and a DDPG agent learns the optimal dispatching of SOP. An edge-intelligence physics-based DRL method is proposed in \cite{qianEdgeIntelligenceBasedFramework2024} for online scheduling of SOPs integrated with energy storage. A DDPG-based DRL method for voltage control of active distributions networks (ADNs) with multi-terminal SOPs is proposed in \cite{liDeepReinforcementLearningBased2022a}. The method features a post-policy action-projection mechanism to avoid infeasible device setpoints and is aimed at regulating grid voltage fluctuations caused by DERs. A hierarchical DRL scheme is proposed in \cite{jiajieOnlineReactivePower2022} for coordinated online dispatching of SOPs and Static Var Compensators (SVCs) in ADNs. Similarly, a hybrid DQN-SAC method is presented in \cite{yinSequentialReconfigurationUnbalanced2023a} for optimal sequential reconfiguration of ADNs with SOPs. Moreover, to manage the distributed topology of active feeders and control devices, Multi-Agent DRL (MARL) architectures deployed under the Centralized Training with Decentralized Execution (CTDE) framework have also been developed for Volt-VAR \cite{chenPhysicsShieldedMultiAgentDeep2023} and SOP dispatching applications \cite{wangDRLBasedActiveReactive2023a}. Virtually all of the foundational actor-critic architectures employed in power distribution applications optimize the scalar expectation of cumulative discounted returns which makes them susceptible to failure modes under asymmetric grid uncertainties.       

A fundamental limitation of DRL applications in power systems is their reliance on critics that track only the expected discounted returns ($\mathbb{E}[Q]$) which fails to account for the stochastic tail-event risks \cite{bellemareDistributionalPerspectiveReinforcement2017} inherent in distribution networks. ADNs experience unexpected operational volatility driven both by weather patterns and customer behavior \cite{liuBiLevelOffPolicyReinforcement2023a}. Robust DRL formulations reported in literature either optimize the policy against worst-case state-perturbations within the considered uncertainty sets \cite{dingActiveExplorationChanceConstrained2022a}, or handle uncertainties by explicitly quantifying uncertainty sets with probabilistic forecasting methods \cite{chenRobustDeepReinforcement2025}. In both cases, the policy update still relies on a scalar Bellman expectation operator. The scalar expectation collapses the underlying multi-modal distribution of returns into a single metric. Distributional DRL (Dist-DRL) entails modeling the full probability of returns using architectures such as quantile regression critics \cite{dabneyDistributionalReinforcementLearning2017} that capture uncertainty skewness and tail-risk boundaries. A search on Dist-DRL applications in power system returns just three contributions; the works presented in \cite{wangMultiagentDistributionalReinforcement2026} and \cite{wangDistributionalReinforcementLearning2026} entail energy management in microgrids while \cite{zhangRiskAwarePowerSystem2026} proposes a method for risk-aware economic dispatch in distribution grids. Despite the distinct benefits of Dist-DRL in safety-critical power distribution applications, the exploration of distributional architectures remains an open research gap.                 

Standard DRL algorithms approximate the process state transition model through repeated exploratory interactions with the environment \cite{guanModelBasedSafeReinforcement2025, chenPhysicsShieldedMultiAgentDeep2023}. When trained for deployment in safety-critical power system applications, the agents lack constraint-handling mechanisms to avoid unsafe regions. Conventional model-free DRL methods also cannot guarantee that predicted control actions satisfy operational safety limits at inference. The limitations of DRL in real-world applications led to the development of a wide variety of Safe Reinforcement Learning (SRL) methods that have risen to prominence in the domain. 

SRL can be broadly categorized into model-free and model-based methods. The most ubiquitous SRL approach models grid operation as a Constrained Markov Decision Process (CMDP). The problem is solved via Lagrangian relaxation \cite{wangSafeOffPolicyDeep2020} in which grid constraint violations are penalized using dynamic Lagrange multipliers. Without extensive tuning, Lagrangian methods exhibit primal-dual oscillations and frequently generate unsafe actions because they only guarantee safety asymptotically in converged policies. In contrast, model-free Trust-Region algorithms such as CPO \cite{achiamConstrainedPolicyOptimization2017} ensure policy updates within the derived safety boundaries but embed a gradient-derived convex optimization loop that inhibits rapid multi-agent dispatch. Physics-informed reinforcement learning (PIRL) and Control Barrier Functions (CBFs) \cite{banerjeeSurveyPhysicsInformed2023} are the two major classes of model-based SRL methods. CBFs often impose overly conservative policy updates that interfere with exploration, and PIRL relies on similar physical modeling dependencies as classical optimization methods. More importantly, current SRL methods in power systems literature almost exclusively ensure safety in converged policies and offer no protection against active exploratory violations in the training phase. A survey of prior art on SRL applications for SOPs reveals little beyond conventional DRL and Lagrangian methods.

Beyond control methodologies, the DRL-based dispatch of SOPs is characterized by a dual-layer constraint challenge that delineates the device-level physical boundaries and the external network operational limits \cite{liDeepReinforcementLearningBased2022a}. At the device-level, the PCSOP dispatch must adhere to (i) the nonlinear apparent power capacity limits across each converter station, and (ii) the internal power loss-coupled strict active power conservation equality constraints across all cross-phase terminals \cite{louMultiterminalPhasechangingSoft2022}. Existing agent-based approaches for SOPs predominantly handle constraint tracking by employing representation relaxations, or by utilizing ad-hoc action clipping and dynamic heuristic masking \cite{liDeepReinforcementLearningBased2022a}. However, device constraint relaxations permit infeasible set-points that might damage the hardware. Conversely, heuristic action-masking distorts the policy gradients, leading to suboptimal policies, while the agent fails to learn to navigate the nonlinear device constraints in unfamiliar instances. These methods leave the agent burdened with learning both the internal action constraints and the complex application objectives. The literature presents no solutions to resolving fully differentiable device-level feasibility for PCSOPs. A structurally constraint-compliant action projection mechanism frees the DRL framework to focus exclusively on grid dispatch optimization. 

A review of the state-of-the-art on SOP considerations reveals a common thread binding the research gaps in the above-mentioned works \cite{liDdpgBasedAdaptiveVoltage2021, yinSequentialReconfigurationUnbalanced2023, xiongReactivePowerOptimization2024, qianEdgeIntelligenceBasedFramework2024, liDeepReinforcementLearningBased2022a, jiajieOnlineReactivePower2022, wangDRLBasedActiveReactive2023a}:
\begin{itemize}
    \item Using device constraint relaxations or disjoint heuristic action-clipping mechanisms to track SOP dispatch boundaries; 
    \item Conflating device constraints and policy learning that forces the agent to learn the nonlinear device physics alongside application objectives through  soft penalties;
    \item Considering two- or multi-terminal SOPs with identical phase sequences and less intricate constraints instead of PCSOPs;
    \item Relying on the scalar Bellman expectation operator for tracking discounted returns that fails to account for operational volatility and stochastic tail-event risks in safety-critical applications;
    \item Using conventional DRL that completely lacks safety nets or primal-dual SRL methods that offer asymptotic-only safety guarantees in expectation while still permitting frequent grid violations throughout training;
    \item Using conventional DRL (DDPG, TD3PG, SAC, PPO) requiring thousands of training episodes to converge;
    \item Running on slow dispatch granularity where DRL methods are predominantly deployed on hour-to-15-minute dispatch cycles to avoid the curse of dimensionality and training instability linked to vanishing temporal gradients;
    \item Not validating methods on a separate test dataset or providing performance comparisons against ACOPF.
\end{itemize}

In general, designing policy learning methods for SOPs is challenging due to interdependencies in their action space. The challenge is compounded in real-world grid imbalance mitigation applications facing increased operational coupling with environmental factors. These problems demand safety guarantees in tightened operational regimes and faster response dynamics to address sub-five-minute DER-induced volatility.

To resolve all seven interconnected gaps previously identified in the data-driven control of SOPs, this paper proposes the Multi-Agent Distributional Wide-Margin Policy Gradient (MADWMPG) framework. MADWMPG is a dual-policy, hybrid Imitation Learning-SRL framework designed for continuous dispatching of distributed, heterogeneous PCSOPs to mitigate grid-wide three-phase voltage unbalance and power flow imbalance at the root node. Operating within the CTDE architecture, the framework guarantees rapidly learned instantaneous constraint safety using decentralized recourse policies and a centralized wide-margin classifier that predicts state-transition feasibility during training. The framework explicitly addresses variability in the demand and DER injection profiles using distributional critics. To orient the policies and accelerate convergence, MADWMPG leverages a strictly offline dataset of static expert demonstrations. This circumvents the impractical requirement of interactive expert querying during training, thus enhancing viability for real-world deployment. The performance of the framework is evaluated on the mixed-phase unbalanced IEEE 123-bus distribution test system. The major contributions of this paper are summarized as follows:
\begin{itemize}
    \item This paper formulates the first recourse-as-a-classification problem in power systems literature. The state-transition classifier and the decentralized recourse policies together form the Wide-Margin Constraint Barrier Recourse (WCBR) module, enabling the PCSOP agents to proactively suppress behavioral policy violations under arbitrary exploratory pressure.  
    \item This paper proposes a distributional target-value generation method for trajectory fidelity as a novel adaptation of Non-crossing Quantile Regression (NQR) critics. The formulation prevents value estimation hallucinations arising from action distribution divergence between the behavioral and recourse policy sets. The distributional critic tracks the full inverse cumulative density function (CDF) of discounted returns, enhancing agent robustness to operational volatility and tail risks. As a supporting contribution, we introduce a recency-biased experience-relay buffer that helps the agent adapt rapidly to policy violations.     
    \item This paper introduces novel hybrid policy loss functions that leverage distributional value learning, the WCBR module, and imitation learning (IL) to enable exploratory safety and ensure convergence to safe inference policies. 
    \item This paper also proposes the first physics-based, fully differentiable zero-shot action projection mechanism for PCSOPs. Structurally embedded within the policy neural networks, the mechanism does not rely on learning processes or heuristic action masking to comply with nonlinear device constraints. The end-to-end differentiable nature of the action sampling mechanism facilitates auto-corrections without distorting the policy gradients. The formulation is extensible and easily adaptable to an arbitrary number of SOP terminals and phase crossings.
    \item The framework is supported by derived probabilistic soft-guarantees on operational safety.  
\end{itemize}        

The framework is trained and tested on an unbalanced multiphase IEEE 123-bus test feeder. The performance is compared to state-of-the art conventional DRL, SRL, and ACOPF methods on an extended operational test dataset. The rest of the paper is organized as follows. Section II describes the overall framework, including the policy networks, recourse policies, the wide-margin classifier and the WCBR module. It also presents the distributional NQR critics, the hybrid policy loss functions, and the probabilistic safety guarantees that bound the performance of the framework. Section III presents the training and performance of the proposed framework, and Section IV concludes the paper.      

\section{THE PROPOSED MADWMPG FRAMEWORK}
\subsection{Background and Problem Statement}
The MADWMPG is a hybrid multi-agent SRL framework for coordinated voltage unbalance mitigation in ADNs equipped with distributed PCSOPS. The proposed framework integrates distributional policy-gradient learning, wide-margin safety classification with recourse, and offline expert imitation within a dual-policy architecture. Each PCSOP agent maintains a primary behavior policy that optimizes performance and a recourse policy that generates corrective actions when the behavior policy is predicted to violate grid operating constraints. A centralized wide-margin classifier evaluates the safety of joint agent actions while the centralized NQR critic estimates the distribution of cumulative returns under stochastic load and DER variability. The MADWMPG aims to maximize operating performance while proactively suppressing unsafe control actions during both exploration and inference. The framework decouples SOP device-level feasibility from network-level operational safety. The operational constraints of PCSOPs are incorporated into the policy architecture through a physics-based zero-shot compliance mechanism, while the framework remains primarily responsible for the network operating constraints. This decomposition enables the behavior and recourse policies to optimize the complementary application objectives without requiring reward-shaping to enforce physical operating limits. 

The voltage and power flow imbalance mitigation problem is modeled as a finite-horizon constrained Markov decision process (CMDP). The CMDP is defined by the tuple $\mathcal M=
\left\{\mathcal S,\mathcal A,\mathsf P,r,c,\nu_0,\gamma,T\right\}$, where \(\mathcal S\) and \(\mathcal A\) denote the global state and joint action spaces respectively, \(\mathsf P\) denotes the state-transition probability function, \(r\) and \(c\) represent the operational reward and safety-cost functions, \(\nu_0\) is the initial-state distribution, \(\gamma\in[0,1]\) is the discount factor, and $T$ is the length of the operating horizon. The time-indexed state transition probabilities are characterized by uncertain load and DER injections, whose prediction intervals are supplied externally to the environment. The stochastic transitions allow the agents to learn policies that account for variations in network operating conditions. Consider a three-phase unbalanced ADN with three multi-terminal PCSOPs. The control objective is to mitigate phase imbalance under stochastic load and DER conditions while satisfying the network voltage limits and the PCSOP dispatch boundaries. At time \(t\), the agents observe the state $\mathbf{s}_t$ comprising the net active $\boldsymbol{p}_t$ and reactive power $\boldsymbol{q}_t$ injections across the grid. Each agent \(n\in\mathcal N=\{1,2,3\}\) maintains a primary behavior policy \(\mu_n\) and a recourse policy \(\rho_n\), each parameterized to generate the nominal and projected PCSOP actions. The state, and the behavior and recourse policy actions are defined by:
\begin{equation}
\label{eq:1}
\begin{aligned}
&\mathbf{s}_t=
\begin{bmatrix}
\boldsymbol{p}_t^\top &
\boldsymbol{q}_t^\top
\end{bmatrix} \\
&\mathbf a_{n,t}^{b}=\mu_n(\mathbf s_t)\\
&\mathbf a_{n,t}^{r}=\rho_n(\mathbf s_t,\mathbf a_{n,t}^{b})
\end{aligned}
\end{equation}
where $a_{n,t}^{b}$ and $a_{n,t}^{r}$ represent the behavior and recourse actions, respectively. The centralized safety classifier determines which joint action is executed. Three-phase voltage unbalance is conventionally quantified by the negative and zero-sequence voltage components, which capture deviations from balanced operation, whereas the deviations from the nominal voltage typically define voltage regulation. The voltage unbalance and regulation objective in (\ref{eq:2}) penalizes both deviations from nominal magnitude and voltage asymmetry:
\begin{equation}
\label{eq:2}
\begin{aligned}
\mathcal L_{\mathrm{v,seq},t} &= \mathcal L_{\mathrm{seq},t}+\mathcal L_{\mathrm{v},t}\\
=\sum_{i\in\Omega_{B,3}}(|V_{0,i,t}|^2+|V_{2,i,t}|&^2)+\sum_{i\in\Omega_B,\phi\in\Omega_{\phi,i}}(1-v_{i,t}^{\phi})^2
\end{aligned}
\end{equation}
where \(\Omega_{B,3}\) denotes the set of three-phase buses, \(\Omega_B\) is the set of all network buses, \(\Omega_{\phi,i}\) contains the phases present at bus \(i\), \(V_{0,i,t}\) and \(V_{2,i,t}\) are the zero- and negative-sequence voltage components, and \(v_{i,t}^{\phi}\) is the corresponding voltage magnitude. The voltage magnitude deviations are calculated for all energized buses whereas the sequence components are restricted to three-phase buses where the transformation permits a meaningful evaluation. The sequence components are nonlinear functions of phase-voltage magnitudes and angles and are considerably smaller than the postive sequence component under typical network operating conditions. Using (\ref{eq:2}) directly as reward penalties produces weak variations in the optimization signal and leads to the agent favoring the dominant voltage regulation metric. To address these limitations, the framework proposes a functionally equivalent reward objective that explicitly exposes the agent to both the angular and voltage magnitude deviations. The desirable operating point is characterized by grid voltage maintained near 1 p.u. and the phase angles approaching positive sequence rotation. Therefore, MADWMPG employs the following voltage objective:
\begin{equation}
\label{eq:3}
\begin{aligned}
\mathcal L_{\mathrm{v,\theta},t} &= \mathcal L_{\theta,t}+\mathcal L_{\mathrm{v},t}\\
=\sum_{i\in\Omega_{B,3}}\Big[(\theta_{i,t}^{a})^2+(\theta_{i,t}^{b}&+2\pi/3)^2+(\theta_{i,t}^{c}-2\pi/3)^2\Big]\\
+\sum_{i\in\Omega_B,\phi\in\Omega_{\phi,i}}&(1-v_{i,t}^{\phi})^2
\end{aligned}
\end{equation}
where \(\theta_{i,t}^{\phi}\) denote the voltage phase angles. The angular penalties act as a quadratic surrogate for deviations from the reference configuration while providing a direct, differentiable metric for optimization. While the voltage objective aims to suppress undesirable sequence components, it does not reduce loading asymmetry at the feeder head. Since uneven power flow at the root node contributes to disproportionate phase loading, the framework incorporates a secondary feeder-head apparent-power imbalance objective as follows:         
\begin{equation}
\label{eq:4}
\mathcal L_{s,t}=(s_{0,t}^{a}-s_{0,t}^{b})^2+(s_{0,t}^{b}-s_{0,t}^{c})^2+(s_{0,t}^{c}-s_{0,t}^{a})^2
\end{equation}
where \(s_{0,t}^{\phi}\) is the feeder-head apparent-power magnitude of phase \(\phi\). Minimizing the objective complements the voltage-oriented measures and promotes a more balanced distribution of feeder loading. The quadratic objectives exhibit considerable differences in range and are therefore integrated into a bounded reinforcement learning reward using exponential transformations:   
\begin{equation}
\label{eq:5}
r_t=\frac{1}{3}\Big[e^{-k_s\mathcal L_{s,t}}
+e^{-k_v\mathcal L_{v,t}}+e^{-k_\theta\mathcal L_{\theta,t}}\Big]
\end{equation}
where \(k_s,k_v,k_\theta>0\) determine the sensitivity of each reward component. The formulation prevents large individual penalties from dominating the reward signal while jointly promoting angular symmetry, voltage regulation, and feeder-head power balancing. Each component yields a theoretical maximum value of 1.0 when its associated penalty vanishes.   

\begin{figure}[!t]
  \centering
  \includegraphics[width=1\linewidth]{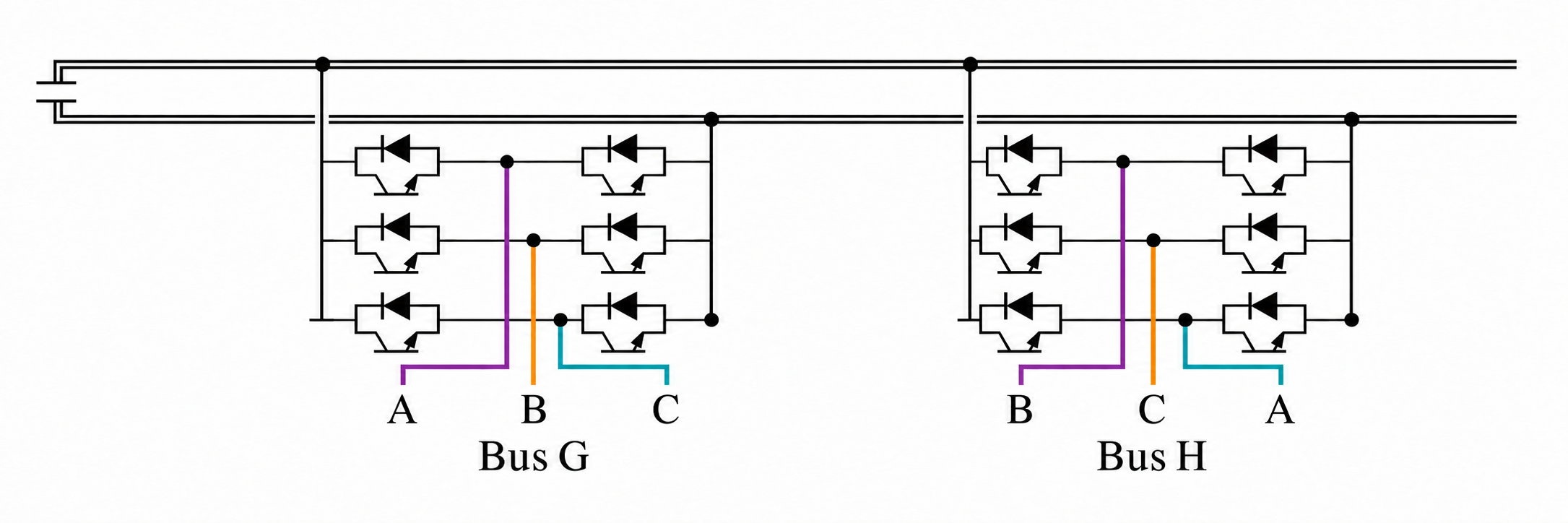}
  \caption{Multi-terminal three-phase PCSOP with color-coded Tie-groups.}
  \label{fig:pcsop}
\end{figure}

The PCSOP actions must satisfy the physical limitations of the device and the distribution network. These include bus voltage limits, the active power equality constraints within each PCSOP tie-group, and the circular converter apparent power capacity constraints:
\begin{equation}
\label{eq:6}
\begin{aligned}
V_{\min}&\le v_{i,t}^{\phi}\le V_{\max}&&\forall i,\phi,t\\
\sum_{(\chi,\phi)\in\mathcal X_{n,g}}&\left(P_{n,\phi,t}^{\chi}+P_{\mathrm{loss},n,\phi,t}^{\chi}\right)=0&&\forall n,g,t\\
(P_{n,\phi,t}^{\chi})^2&+(Q_{n,\phi,t}^{\chi})^2\le S_{n,\phi}^2&&\forall n,\chi,\phi,t
\end{aligned}
\end{equation}
With reference to Fig. \ref{fig:pcsop}, \(\mathcal X_{n,g}\) denotes the set of terminal-phase pairs belonging to tie-group \(g\) of PCSOP \(n\), \(P_{n,\phi,t}^{\chi}\) and \(Q_{n,\phi,t}^{\chi}\) are the corresponding active and reactive power injections, and \(S_{n,\phi}\) is the converter capacity. The nonlinear PCSOP losses are a function of the active-reactive power injections \cite{louNewPhasechangingSoft2020}:
\begin{equation}
\label{eq:7}
P_{\mathrm{loss},n,\phi,t}^{\chi}=\Gamma\sqrt{(P_{n,\phi,t}^{\chi})^2+(Q_{n,\phi,t}^{\chi})^2}
\end{equation}
where \(\Gamma\) is a device parameter. The PCSOP power-balance and capacity constraints are device-level limitations that can be incorporated into the physics-based action generation mechanism. In contrast, the bus voltages depend on the network's response to the collective actions of the PCSOP agents. Maintaining voltage security requires joint-action assessment rather than independent action corrections. The instantaneous grid-wide voltage-violation cost is defined as:
\begin{equation}
\label{eq:8}
\begin{aligned}
c_t=\sum_{i\in\Omega_B}\sum_{\phi\in\Omega_{\phi,i}}&\mathrm{max}\Big(0,|V_{\min}-v_{i,t}^{\phi}|\Big)\\
&+\mathrm{max}\Big(0,|v_{i,t}^{\phi}-V_{\max}|\Big)
\end{aligned}
\end{equation}
The nonnegative cost \(c_t\) is zero when all phase voltages are within the desired limits and increases with the magnitude of violations. The unchecked execution of behavior policy actions, particularly during exploration, can lead to constraint violations. MADWMPG introduces a centralized wide-margin safety classifier which evaluates the joint action proposals and selects between the behavior and recourse policies:
\begin{equation}
\label{eq:9}
\begin{aligned}
I_t&=\mathbb I\!
\left[C_\omega(\mathbf s_t,\mathbf a_t^b)>0\right]\\
\mathbf a_t^g
&=I_t\mathbf a_t^b+(1-I_t)\mathbf a_t^r
\end{aligned}
\end{equation}
where \(C_\omega\) is the learned safety-classification function, \(I_t\) is the binary gating indicator, and \(\mathbf a_t^g\) is the action executed in the environment. The gating mechanism permits performance-oriented exploration when deemed safe and falls back to a safer alternative otherwise. Thus, to obtain the set of optimal policies, the proposed framework pursues complementary objectives of maximizing the cumulative application reward and the classifier-assessed safety scores of both the behavior and recourse policies. The optimal policies are defined as:
\begin{equation}
\label{eq:10}
\begin{aligned}
\max_{\boldsymbol\mu,\boldsymbol\rho}\mathbb E_\tau\!\left[\sum_{t=0}^{T-1}\gamma^t\big(r_t+C_\omega(\mathbf s_t,\mathbf a_t^b)+C_\omega(\mathbf s_t,\mathbf a_t^r)\big)\right]
\end{aligned}
\end{equation}
where \(\tau\) is the trajectory of state-action pairs over the finite operating horizon of \(T\) steps. 

\subsection{The Expert Demonstrator, Parallelized Stochastic Environments, and Recency-biased Experience Replay}
The framework leverages IL using a set of expert-defined state-action pairs to orient the agents and accelerate policy convergence. The expert demonstrator is a nonlinear ACOPF program that solves for the objective defined in (\ref{eq:5}) over an operating horizon spanning a day's operation. The program considers the multi-phase nature of the unbalanced grid, the complex power flow in mutually coupled line phases, and the voltage and nonlinear PCSOP constraints. The ACOPF formulates a nonlinear and nonconvex problem that often stalls or fails to converge without proper constraining of the solution space. To improve tractability, the program is augmented with an auxilary set of constraints related to the reduction of apparent power flow imbalance at the feeder-head:
\begin{equation}
\label{eq:11}
\sum_{t=0}^{T-1}\Big(|s_{0,t}^{a}-s_{0,t}^{b}|+|s_{0,t}^{b}-s_{0,t}^{c}|+|s_{0,t}^{c}-s_{0,t}^{a}|\Big)\le\delta_s
\end{equation}
where \(\delta_s\) is the mismatch tolerance. The ACOPF therefore solves the following program:
\begin{equation}
\label{eq:12}
\begin{aligned}
\max &\sum_{t=0}^{T-1} r_t \\
\mathrm{subject \ to}: \ &\mathrm{(6), \ (7), \ (11)}
\end{aligned}
\end{equation}
The state-action pairs obtained from the solution are saved in a separate buffer, \(\mathcal D_{\mathrm{ex}}\), and used for policy updates.

To expose the agents to operational variability and improve data density, the training environment is parallelized across \(N_E\) power flow routines. The power flow routine is similar to the ACOPF except that it treats the agents' actions as parameters. Each instance of the environment represents an independent realization of the stochastic temporal load demand and DER power injections. The uncertainty in load and generation profiles is introduced through externally supplied time-varying prediction intervals. Each uncertain quantity \(\zeta\) is generated using uniform sampling within its prescribed bounds.      
\begin{equation}
\label{eq:13}
\begin{aligned}
\zeta_{e,j,t+1}
&\sim\mathcal U\left(
\underline{\zeta}_{e,j,t+1},
\overline{\zeta}_{e,j,t+1}
\right),\\
\mathbf s_{e,t+1}
&=\mathcal H(\boldsymbol\zeta_{e,t+1}),
\end{aligned}
\end{equation}
where \(e\in\{1,\ldots,N_E\}\) indexes the parallel environments, \(j\) identifies an uncertain load or generation quantity, and \(\mathcal H\) maps the sampled quantity to the net active and reactive power injections. Uniform sampling does not require a specific probability density model for the underlying uncertainty. At each training iteration, the agents interact with every parallel environment using the joint actions filtered by the safety-gating mechanism in (\ref{eq:9}), and the resulting process transitions \(\xi\) are stored in the replay buffer.    
\begin{equation}
\label{eq:14}
\xi_{e,t}=\left(\mathbf s_{e,t},\mathbf a_{e,t}^{g},r_{e,t},\mathbf s_{e,t+1},y_{e,t}\right)
\end{equation}
where \(y_{e,t}\in\{-1,+1\}\) is the binary safety label assigned according to the observed voltage-violation cost in (\ref{eq:8}). 
\begin{equation}
\label{eq:15}
y_{e,t}=
\begin{cases}
+1, & c_{e,t}=0,\\
-1, & c_{e,t}>0.
\end{cases}
\end{equation}

The random sampling of experiences in conventional experience replay underrepresents recent state visitations, particularly as the replay buffer grows. Changes in stochastic operating conditions and newly encountered violations provide valuable feedback for performance and safety corrections. Consequently, the safety classifier and the critic must adapt to recent trajectories to accurately pivot between behavioral and recourse actions. To address this, experience replay is augmented with a recency-biased sampling mechanism that combines uniformly sampled experiences with the most recent \(k\) transitions generated by each parallel environment.    
\begin{equation}
\label{eq:16}
\begin{aligned}
\widetilde{\mathcal B}_t
&=\mathcal B_t\Vert\mathcal B_t^{\mathrm{rec}},\\
|\widetilde{\mathcal B}_t|
&=|\mathcal B_t|+kN_E.
\end{aligned}
\end{equation}
where \(\mathcal B_t\) is the minibatch of uniformly sampled experiences, \(\mathcal B_t^{\mathrm{rec}}\) are the recent transitions, and \(\Vert\) denotes concatenation. \(\mathcal B_t^{\mathrm{rec}}\) contains \(kN_E\) samples. The mechanism enables broader reuse of historical samples with greater emphasis on current operating conditions.   

\subsection{Physics-Based Zero-Shot PCSOP Action Generation}
The active and reactive power injections of a PCSOP must satisfy the per-phase apparent power capacity and the loss-coupled active-power conservation constraint within each phase tie-group. Enforcing these constraints through reward penalties alone forces agents to learn the underlying physical relationships. An external action-projection mechanism adds optimization overhead and breaks the gradient backpropagation chain. The MADWMPG framework avoids these limitations using a physics-based zero-shot predictor-corrector mechanism that combines capacity-aware action generation with differentiable corrective redispatch. 
\subsubsection{Action Prediction}

Consider the two-terminal PCSOP connecting buses G and H in Fig. \ref{fig:pcsop}, comprising three phase tie-groups $(G_a,H_b)$, $(G_b,H_c)$, and $(G_c,H_a)$. The policy generates the reactive power commands for both terminals but the active power command for only one, the remaining active power component is determined by the power balance equation. The independent action components are:
\begin{equation}
\label{eq:17}
\begin{aligned}
\mathbf Q^{\mathrm{ind}}
&=[Q_G^a,Q_G^b,Q_G^c,
Q_H^b,Q_H^c,Q_H^a]^\top\\
\mathbf P^{\mathrm{ind}}
&=[P_G^a,P_G^b,P_G^c]^\top
\end{aligned}
\end{equation}
This parametrization reduces the variable count, allowing the remaining injections to be reconstructed directly from the PCSOP physics for both behavior and recourse policies. To produce capacity-compliant independent actions, let $\mathbf h_n$ denote the latent representation of agent $n$, with reactive and active power output heads $f_{Q,n}$ and $f_{P,n}$. The normalized policy outputs are: 
\begin{equation}
\label{eq:18}
\begin{aligned}
\mathbf q_n^{(0)}
&=\tanh\!\left(f_{Q,n}(\mathbf h_n)\right)\\
p_{n,j}^{(0)}
&=\tanh\!\left(f_{P,n}(\mathbf h_n)_j\right)
\sqrt{1-\left(q_{n,j}^{(0)}\right)^2}
\end{aligned}
\end{equation}
The hyperbolic tangent bounds reactive power to 1 pu. Multiplying the active power output by the available capacity $\sqrt{1-q_j^2}$ ensures the dispatch satisfies the terminal's apparent power capacity constraint, preventing excessive active power requests during high reactive power utilization. During training, Gaussian perturbations $\epsilon_j^p$ and $\epsilon_j^q$ are introduced. Altering the reactive power changes the available active power capacity and the limit must be recomputed before dynamic clipping of the active power components. 
\begin{equation}
\label{eq:19}
\begin{aligned}
q_j^e
&=\operatorname{clip}
(q_j^{(0)}+\epsilon_j^q,-1,1)\\
p_{j,\max}^e
&=\sqrt{1-(q_j^e)^2}\\
p_j^e
&=\operatorname{clip}
(p_j^{(0)}+\epsilon_j^p,
-p_{j,\max}^e,p_{j,\max}^e)
\end{aligned}
\end{equation}
This maintains capacity compatibility at the independent terminal. However, the coupled dependent terminal must still analytically satisfy physical balance and losses. Consider a representative tie-group with normalized injections $(p_1,q_1)$ and $(p_2,q_2)$ corresponding to phases \(G_a\) and \(H_b\). With $p_2$ as the sole unknown, active-power balance requires:
\begin{equation}
\label{eq:20}
\begin{aligned}
&p_1+p_2+\Gamma\Big(\sqrt{p_1^2+q_1^2}+\sqrt{p_2^2+q_2^2}\Big)=0\\
&\kappa=-p_1-\Gamma\sqrt{p_1^2+q_1^2}
\end{aligned}
\end{equation}
The quantity $\kappa$ isolates the independent terminal's contributions. The unknown term \(p_2\) is calculated by transforming (\ref{eq:20}) into a quadratic equation:
\begin{equation}
\label{eq:21}
\begin{aligned}
Ap_2^2+Bp_2+C&=0\\
p_2^\pm&=\frac{-B\pm\sqrt{\Delta}}{2A}
\end{aligned}
\end{equation}
where $A=\Gamma^2-1$, $B=2\kappa$, $C=\Gamma^2q_2^2-\kappa^2$, and $\Delta=B^2-4AC$. The solution roots may not satisfy the original equality or exceed the terminal's apparent power rating. The solutions are therefore checked against both the power balance equations and the capacity constraints:
\begin{equation}
\label{eq:22}
\begin{aligned}
e_\sigma={}&\Big|p_1+p_2^\sigma+\Gamma\Big(\sqrt{p_1^2+q_1^2}\\
&\qquad+\sqrt{(p_2^\sigma)^2+q_2^2}\Big)\Big|\\
m_\sigma={}&\mathbb I\Big\{e_\sigma\le\varepsilon_b,\;(p_2^\sigma)^2+q_2^2\le1\Big\}
\end{aligned}
\end{equation}
where $\sigma\in\{+,-\}$, $e_\sigma$ is the residual, and $\varepsilon_b=10^{-3}$ p.u. is the tolerance. The positive-branch root is preferred if valid; otherwise, the negative branch is considered. If neither passes, the corrective redispatch routine is initiated.

\subsubsection{Corrective Redispatch of Infeasible Actions}
An infeasible result in (\ref{eq:21}) implies the dependent variable cannot accommodate the transfer requirement. The corrector routine first modifies the independent active-power request using the maximum absolute discrepancy $\delta$ between the algebraic candidates and the capacity boundary:
\begin{equation}
\label{eq:23}
\begin{aligned}
p_{2,\max}&=\sqrt{1-q_2^2}\\
\delta&=\max_{\sigma\in\{+,-\}}\left|p_{2,\max}-|p_2^\sigma|\right|\\
p_1'&=p_1-\delta\operatorname{sgn}(p_1)\\
q_1'&=q_1\\
\kappa'&=-p_1'-\Gamma\sqrt{(p_1')^2+(q_1')^2}
\end{aligned}
\end{equation}
The discrepency $\delta$ reduces the independent active-power request $p_1'$ in the direction opposite to its original sign thereby decreasing the transfer burden on the coupled terminal. The revised right-hand side $\kappa'$ incorporates this updated injection and its corresponding loss. If the dependent terminal already uses substantial capacity for reactive power, reducing the active-power request may be insufficient. The corrector then adjusts $q_2$ based on a fictitious case where dependent reactive power is zero:
\begin{equation}
\label{eq:24}
\begin{aligned}
p_{2,0}
&=
\begin{cases}
\kappa'/(1+\Gamma),&\kappa'>0,\\
\kappa'/(1-\Gamma),&\kappa'\le0,
\end{cases}\\
q_2'&=0.99q_2
\sqrt{\max\{0,1-p_{2,0}^2\}}
\end{aligned}
\end{equation}
The reference $p_{2,0}$ estimates the active-power requirement before reactive-power loading is reintroduced. Scaling the original reactive command by the remaining apparent-power headroom dynamically suppresses reactive-power utilization as the active-power limit is approached. The $0.99$ factor ensures a numerical safety margin. The active and reactive power adjustments change $\kappa'$ and the capacity circle, the original quadratic roots are no longer valid. The corrector reconstructs the quadratic equation:
\begin{equation}
\label{eq:25}
\begin{aligned}
p_2^{\prime\pm}=\frac{-B'\pm\sqrt{\max\{0,\Delta'\}}}{2A}
\end{aligned}
\end{equation}
where $B'=2\kappa'$, $C'=\Gamma^2(q_2')^2-(\kappa')^2$, and $\Delta'=(B')^2-4AC'$. These corrected candidates are evaluated against the unsquared balance equation and capacity constraints using $(p_1',q_1',q_2')$. The positive-branch candidate is preferred when valid, with the alternative branch considered otherwise. 

In summary, the piecewise differentiable predictor–corrector generates capacity-compatible actions, analytically computes dependent injections, and redispatches infeasible actions. The explicit nonlinear transformations and closed-form calculations enable zero-shot physical compliance without iterative optimization or independent feasibility models, leaving network-wide voltage safety strictly to the centralized wide-margin classifier. 

\subsection{The Wide-Margin Constraint Barrier Recourse Module}
The physics-based action generation mechanism in section II-C ensures device-level compliance but cannot guarantee network constraint satisfaction. The bus voltages depend on the network conditions and the combined actions of all PCSOP agents. Consequently, the device-compliant actions may still produce grid-wide violations. MADWMPG addresses this through a centralized wide-margin safety classifier and learned recourse policies. The centralized state-transition classifier \(C_\omega(\mathbf s_t,\mathbf a_t)\) estimates the safety of a joint PCSOP action by evaluating the state $\mathbf s_t$ and the concatenated actions of all agents to capture the coupled network response. The sign of the classifier's score dictates the predicted safety class, and its magnitude indicates the degree of separation from the classification boundary. The classifier is trained using binary labels \(y_t\in\{-1,+1\}\) assigned by observing the network voltage response. To encourage maximal separation, the framework adopts a support-vector-machine-inspired wide-margin hinge loss:
\begin{equation}
\label{eq:26}
\begin{aligned}
C_t&=C_\omega(\mathbf s_t,\mathbf a_t^g)\\
\ell_{\mathrm{hinge}}(y_t,C_t)
&=\max(0,1-y_tC_t)
\end{aligned}
\end{equation}
The loss penalizes both misclassifications (\(y_tC_t\le0\)) and correct classifications that fall close to the unit margin (\(0<y_tC_t<1\)). The continuous learning signal pushes actions away from the classifier boundary, which is critical near operating limits where small variations can change safety labels. The training objective using the recency-biased minibatch is:
\begin{equation}
\label{eq:27}
\begin{aligned}
\mathcal L_C(\omega)=
\frac{1}{|\widetilde{\mathcal B}_t|}\sum_{\xi_j\in\widetilde{\mathcal B}_t}\Big[1-y_j C_\omega(\mathbf s_j,\mathbf a_j^g)\Big]_+
\end{aligned}
\end{equation}
Importantly, the classifier training uses the executed joint actions \(\mathbf a_j^g\) and its observed safety label rather than the unexecuted behavior or recourse actions. This guarantees every label corresponds to an actual power flow evaluation. Rejecting a potentially unsafe action requires an alternative. The proposed framework therefore associates each behavior policy \(\mu_n\) with a recourse policy \(\rho_n\), which learns a state-dependent remapping function that is conditioned on the behavior action:      
\begin{equation}
\label{eq:28}
\mathbf a_{n,t}^{r}=\rho_n\left(\mathbf s_t,\mathbf a_{n,t}^{b};\boldsymbol\theta_n^\rho\right)
\end{equation}
where \(\boldsymbol\theta_n^\rho\) denotes the parameters of recourse policy \(n\). Action-conditioning allows identical states to produce different recourse actions depending on which PCSOP action contributes to the violation. The recourse policies are trained using both the distributional critic for expected return and the safety classifier for predicted safety, which is assigned a greater weight. As shown in (\ref{eq:9}), the centralized classifier assesses the joint behavior action proposal. If predicted safe, the action is passed to the environment; otherwise, the recourse action is selected. The indicator \(I_t\) in (\ref{eq:9}) operates on the complete joint behavior action to prevent the execution of independently accepted but mutually incompatible actions. The safety classifier operates in a loop where optimizing the recourse policies generates new transitions and safety labels which in turn update the classifier and refine the safety landscape. The interactions are supported by the recency-biased experience replay. It should be noted that the extent to which the residual safety violations affect the closed-loop safety is addressed by the probabilistic bounds derived in section II-F. Thus, the framework implements a hierarchical architecture: the zero-shot action generator addresses the device-level action relationships while WCBR addresses the grid-wide safety consequences.               

\subsection{Noncrossing Distributional Critics and Hybrid Dual-Policy Optimization}
Conventional expectation-tracking methods fail to capture return variability from stochastic operating conditions and their nonlinear influence on the distribution network. The proposed framework addresses this using a centralized NQR distributional critic that estimates the complete return distribution rather than a single expected value. For a trajectory of length \(T\) with discount factor \(\gamma\in[0,1)\), the discounted return is:
\begin{equation}
\label{eq:29}
Z_t=\sum_{h=0}^{T-t-1}\gamma^h r_{t+h}
\end{equation}
Instead of a parametric density, the centralized critic \(Z_\psi(\mathbf s,\mathbf a)\), parameterized by \(\psi\), approximates this distribution using $M$ quantile atoms evaluated at uniformly spaced probability fractions. 
\begin{equation}
\label{eq:30}
\begin{aligned}
\tau_m&=\frac{m-\frac12}{M},\quad m=1,\ldots,M\\
\mathbf z_\psi(\mathbf s,\mathbf a)
&=\left[z_\psi^{(1)},\ldots,z_\psi^{(M)}\right]^\top
\end{aligned}
\end{equation}
The critic conditions this distribution on the network state and the concatenated joint PCSOP actions because the reward reflects system-wide imbalance in response to the coordinated actions of all three PCSOP agents rather than any individual action alone. The quantile fractions \(z_\psi^{(m)}\) provide a discrete approximation of the entire conditional return distribution, allowing the critic to discern lower-return outcomes from favorable realizations. To ensure the estimated quantiles form a consistent discrete representation of the return distribution, the NQR critic enforces a noncrossing representation that explicitly maintains monotonicity:
\begin{equation}
\label{eq:31}
z_\psi^{(1)}(\mathbf s,\mathbf a)\le \cdots \le z_\psi^{(M)}(\mathbf s,\mathbf a)
\end{equation}
The critic must be trained consistently with the framework's safety-gated execution mechanism, as using the ungated behavior action proposals in the Bellman target would estimate future returns under an action distribution that diverges from the one used during experience sampling. The target behavior and recourse policies first generate actions at the sampled next state using the recency-biased minibatch:
\begin{equation}
\label{eq:32}
\begin{aligned}
\bar{\mathbf a}_{n,j+1}^{b}&=\bar\mu_n(\mathbf s_{j+1})\\
\bar{\mathbf a}_{n,j+1}^{r}&=\bar\rho_n\left(\mathbf s_{j+1},\bar{\mathbf a}_{n,j+1}^{b}\right)
\end{aligned}
\end{equation}
The centralized classifier evaluates the target behavior action and applies the binary selection mechanism to obtain the gated target actions \(\bar{\mathbf a}_{j+1}^{g}\):
\begin{equation}
\label{eq:33}
\begin{aligned}
\bar I_{j+1}&=\mathbb I\left[C_\omega(\mathbf s_{j+1},\bar{\mathbf a}_{j+1}^{b})>0\right]\\
\bar{\mathbf a}_{j+1}^{g}&=\bar I_{j+1}\bar{\mathbf a}_{j+1}^{b}+(1-\bar I_{j+1})\bar{\mathbf a}_{j+1}^{r}
\end{aligned}
\end{equation}
Using this joint target action, each target quantile \(\mathcal Y_j^{(\ell)}\) incorporates the immediate reward and the corresponding bootstrapped quantile of the target critic \(z_{\bar\psi}\):
\begin{equation}
\label{eq:34}
\begin{aligned}
\mathcal Y_j^{(\ell)}=r_j+\gamma z_{\bar\psi}^{(\ell)}\left(\mathbf s_{j+1},\bar{\mathbf a}_{j+1}^{g}\right)
\end{aligned}
\end{equation}
The critic is trained using quantile Huber regression, which compares every predicted quantile with all $M$ target samples using the temporal distributional residual \(\delta_{j,m\ell}\):
\begin{equation}
\label{eq:35}
\begin{aligned}
\delta_{j,m\ell}=\mathcal Y_j^{(\ell)}-z_\psi^{(m)}\left(\mathbf s_j,\mathbf a_j^g\right).
\end{aligned}
\end{equation}
Quantile regression uses asymmetric weighting according to whether the target lies above or below the predicted quantile. To achieve proper quantile estimation while reducing sensitivity to large errors, the asymmetrically weighted quantile Huber loss is used, where $\kappa>0$:
\begin{equation}
\label{eq:36}
\begin{aligned}
\rho_{\tau_m}^{\kappa}(u)&=\left|\tau_m-\mathbb I(u<0)\right|\frac{\mathcal H_\kappa(u)}{\kappa}\\
\mathcal H_\kappa(u)&=\begin{cases}\frac12u^2,& |u|\le\kappa\\
\kappa\left(|u|-\frac12\kappa\right),& |u|>\kappa
\end{cases}
\end{aligned}
\end{equation}
The critic objective \(L_Z\) averages the pairwise quantile Huber losses over the recency-biased training minibatch $\widetilde{\mathcal B}_t$ samples and is normalized by $M^2$ for pairwise comparisons:
\begin{equation}
\label{eq:37}
\begin{aligned}
\mathcal L_Z(\psi)={}\frac{1}{|\widetilde{\mathcal B}_t|M^2}\sum_{\xi_j\in\widetilde{\mathcal B}_t}\sum_{m=1}^{M}\sum_{\ell=1}^{M}\rho_{\tau_m}^{\kappa}\left(\delta_{j,m\ell}\right)
\end{aligned}
\end{equation}
The required scalar performance estimate for deterministic policy optimization is extracted by averaging over the uniformly spaced return quantiles:
\begin{equation}
\label{eq:38}
\widehat Q_\psi(\mathbf s,\mathbf a)=\frac{1}{M}\sum_{m=1}^{M}z_\psi^{(m)}(\mathbf s,\mathbf a)
\end{equation}
The expectation of the return distribution \(Q_\psi(\mathbf s,\mathbf a)\) and the safety classifier $C_\omega$ jointly optimize the dual policy set. The behavior policies are optimized for grid performance and recourse policies prioritize voltage safety. The behavior policies mitigate network imbalance by maximizing (or minimizing the negative) expected return of their joint action $\mathbf a_j^b=\boldsymbol\mu(\mathbf s_j)$:
\begin{equation}
\label{eq:39}
\mathcal L_{\mu,Q}=-\frac{1}{|\widetilde{\mathcal B}_t|}\sum_{\xi_j\in\widetilde{\mathcal B}_t}\widehat Q_\psi(\mathbf s_j,\mathbf a_j^b)
\end{equation}
To encourage generation of safer behavior proposals and reduce reliance on the recourse mechanism, the behavior policy loss is augmented with classifier guidance that maximizes the predicted continuous safety score of the behavior actions:
\begin{equation}
\label{eq:40}
\mathcal L_{\mu,C}=-\frac{1}{|\widetilde{\mathcal B}_t|}\sum_{\xi_j\in\widetilde{\mathcal B}_t}C_\omega(\mathbf s_j,\mathbf a_j^b)
\end{equation}
Furthermore, to improve sample efficiency without requiring online ACOPF solutions and to orient agents, the behavior policies are also regularized against an offline fixed dataset of expert demonstrations $\mathcal D_{\mathrm{ex}}$. At each update instance, the behavior policies generate actions which are compared to the expert reference \(a^{\mathrm{ex}}\) for the sampled expert states \(\mathbf s^{\mathrm{ex}}\). The total imitation loss using the expert minibatch $\mathcal B_t^{\mathrm{ex}}$ over joint action dimension $d_a$ is:
\begin{equation}
\label{eq:41}
\begin{aligned}
\mathbf a_j^{b,\mathrm{ex}} &=\boldsymbol\mu(\mathbf s_j^{\mathrm{ex}})\\
\mathcal L_{\mathrm{IL}}&={}\frac{1}{|\mathcal B_t^{\mathrm{ex}}|d_a}\sum_{j\in\mathcal B_t^{\mathrm{ex}}}\sum_{\ell=1}^{d_a}h_\beta\left(a_{j,\ell}^{b,\mathrm{ex}}a_{j,\ell}^{\mathrm{ex}}\right)
\end{aligned}
\end{equation}
where \(h_\beta\) is the Huber imitation loss. The three behavior loss components are merged into a performance-oriented hybrid policy objective:
\begin{equation}
\label{eq:42}
\mathcal L_\mu=w_1^b\mathcal L_{\mu,Q}+w_2^b\mathcal L_{\mu,C}+w_3^b\mathcal L_{\mathrm{IL}}
\end{equation}
where the order of the loss factors $w_1^b>w_2^b>w_3^b$ prioritizes expected return maximization with safety guidance over expert imitation. The framework harnesses safe rewarding actions in the vicinity of known local optima to enhance the sample efficiency of the learning process. Conversely, the behavior action-conditioned recourse policies generate alternative joint actions \(\mathbf a^r\) only when the proposed action is classified as unsafe. Training the recourse policies does not include an expert-led imitation loss. Their objective reverses the behavior priorities with $w_1^r<w_2^r$, assigning a larger weight to classifier score to emphasize safety correction while retaining a smaller return weight to prevent excessive grid performance degradation:
\begin{equation}
\label{eq:43}
\begin{aligned}
\mathcal L_\rho
=-\frac{1}{|\widetilde{\mathcal B}_t|}\sum_{\xi_j\in\widetilde{\mathcal B}_t}\Big[w_1^r\widehat Q_\psi(\mathbf s_j,\mathbf a_j^r)+w_2^r C_\omega(\mathbf s_j,\mathbf a_j^r)\Big]
\end{aligned}
\end{equation}
The parameters of the dual-policy set are updated by differentiating their respective hybrid losses through the fixed centralized critic and safety classifier parameters:
\begin{equation}
\label{eq:44}
\begin{aligned}
\boldsymbol\theta_n^\mu
&\leftarrow
\boldsymbol\theta_n^\mu
-\alpha_\mu
\nabla_{\boldsymbol\theta_n^\mu}\mathcal L_\mu\\
\boldsymbol\theta_n^\rho
&\leftarrow
\boldsymbol\theta_n^\rho
-\alpha_\rho
\nabla_{\boldsymbol\theta_n^\rho}\mathcal L_\rho
\end{aligned}
\end{equation}
where \(\alpha_\mu\) and \(\alpha_\rho\) are the respective policy learning rates. The gradients \(\nabla_{\boldsymbol\theta_n^\mu}\) and \(\nabla_{\boldsymbol\theta_n^\rho}\) propagate through the differentiable components of the PCSOP physics-based action generation mechanism, including their nonlinear algebraic transformations wherever their active branches are differentiable. The dual-policy learning process combines stochastic environmental experience through distributional critics, constraint satisfaction through the wide-margin classifier, and reference control through offline expert demonstrations.    

\begin{figure*}[!t]
  \centering
  \includegraphics[width=1\textwidth]{{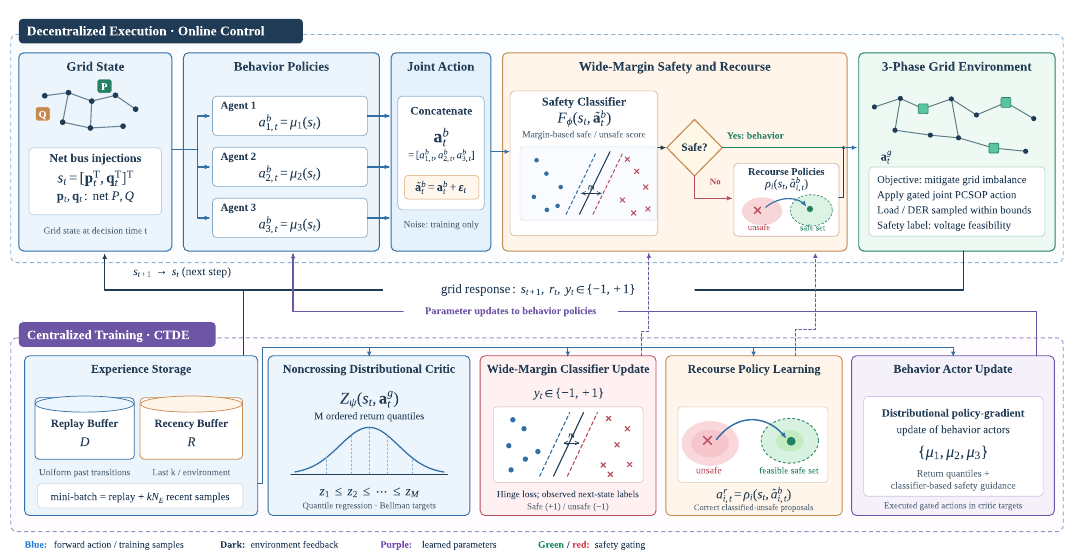}}
  \caption{The MADWMPG framework.}
  \label{fig:madwmpg}
\end{figure*}

\subsection{Probabilistic Guarantees on Safety Performance}
The closed-loop execution safety of the centralized wide-margin classifier and the recourse policies is governed by statistical generalization bounds. To quantify the protection afforded by the MADWMPG framework, this section derives a probabilistic soft guarantee bounding the likelihood that the safety-gated joint action produces network voltage violations during execution. Markov-based bounds on catastrophic failures further support this analysis.

Let $\mathcal{U}$ denote the set of unsafe operating states defined by voltage violations. The true binary safety label $y(\mathbf s, \mathbf a) \in \{+1, -1\}$ indicates whether executing a joint action $\mathbf a$ from state $\mathbf s$ transitions the network int a safe ($+1$) or unsafe ($-1$) condition. Using the framework's gating mechanism, the classifier evaluates only the behavior proposals $\mathbf a^b$. The executed joint action passed to the environment is:
\begin{equation}
\label{eq:45}
\mathbf a^g =
\begin{cases}
\mathbf a^b, & \text{if } C_\omega(\mathbf s, \mathbf a^b) \ge 0 \\ 
\mathbf a^r, & \text{if } C_\omega(\mathbf s, \mathbf a^b) < 0
\end{cases}
\end{equation}
Let $V = \mathbb{I}[\mathbf s' \in \mathcal{U}]$ define a safety violation indicator which equals 1 if the transition is unsafe. The one-step probability of a network violation during execution is the expectation of this indicator, corresponding to the execution failure event $E$:
\begin{equation}
\label{eq:46}
p_{\mathrm{unsafe}}
= \mathbb{E}[V]
= \mathbb{P}\big(y(\mathbf s, \mathbf a^g) = -1\big)
= \mathbb{P}(E)
\end{equation}
Since the classifier only checks $\mathbf a^b$, an execution failure ($E$) occurs through only one of two mutually exclusive paths: the gate incorrectly accepts an unsafe behavior action, or generates a recourse action that is also unsafe following $\mathbf a^b$ rejection. Since the recourse policies are trained to optimize the safety score, the framework operates under the assumption that the proposed alternative actions satisfy $C_\omega(\mathbf s, \mathbf a^r) \ge 0$. Consequently, the executed action $\mathbf a^g$  is always assumed to possess a non-negative predicted safety score with $C_\omega(\mathbf s, \mathbf a^g) \ge 0$. Therefore, any execution failure ($E$) implies:
\begin{equation}
\label{eq:47}
E \implies \left\{ y(\mathbf s, \mathbf a^g) = -1 \text{ and } C_\omega(\mathbf s, \mathbf a^g) \ge 0 \right\}
\end{equation}
To relate $E$ to the classifier's optimization objective, let \(\ell_{0\text{–}1}\) define the 0-1 misclassification loss for the executed action and corresponding set of error events $M$:
\begin{equation}
\label{eq:48}
\begin{aligned}
\ell_{0\text{–}1}\big(y, C_\omega(\mathbf s, \mathbf a^g)\big)
&= \mathbb{I}\big[y(\mathbf s, \mathbf a^g)C_\omega(\mathbf s, \mathbf a^g) \le 0\big]\\
M &= \bigl\{y(\mathbf s, \mathbf a^g)C_\omega(\mathbf s, \mathbf a^g) \le 0\bigr\}
\end{aligned}
\end{equation}
where $M$ denotes the union of all false negative and false positive classification errors. When an event $E$ occurs, it strictly satisfies the condition for misclassification event $M$, establishing the structural inclusion $E \subseteq M$.  Every execution failure maps to a classifier misclassification of $\mathbf a^g$. This implies $\mathbb{I}[E] \le \mathbb{I}[M]$ using the pointwise property of indicator functions for nested sets. Applying the expectation operator converts these indicators to probabilities:
\begin{equation}
\label{eq:49}
\mathbb{E}\big[\mathbb{I}[E]\big]
\le \mathbb{E}\big[\mathbb{I}[M]\big]
\implies p_{\mathrm{unsafe}} \le \mathcal{R}_{0\text{–}1}(\omega)
\end{equation}
where $\mathcal{R}_{0\text{–}1}(\omega)$ is the true $0\text{–}1$ classification risk over the distribution of executed actions. The true hinge risk $\mathcal{R}_{\mathrm{hinge}}(\omega)$ is the expected value of the continuous unit-margin hinge loss over the distribution of executed actions as defined in section II-D and is inclusive of all penalties arising from events in $M$. 
\begin{equation}
\label{eq:50}
\mathcal{R}_{\mathrm{hinge}}(\omega)
= \mathbb{E}\Big[\ell_{\mathrm{hinge}}\big(y(\mathbf s, \mathbf a^g), C_\omega(\mathbf s, \mathbf a^g)\big)\Big]
\end{equation}
Therefore, the violation probability is ultimately bounded by the true hinge risk:     
\begin{equation}
\label{eq:51}
p_{\mathrm{unsafe}}
\le \mathcal{R}_{0\text{–}1}(\omega)
\le \mathcal{R}_{\mathrm{hinge}}(\omega).
\end{equation}
While the expected risk provides a bound on global performance, the continuous safety margin can be analyzed to quantify the frequency of catastrophic failures. Specifically, we derive an upper bound on the probability that an executed action classified as safe by a strict margin $\delta \ge 0$ results in a physical violation. Consider a severe error where an unsafe executed action ($y = -1$) scores a highly safe classifier score ($C_\omega(\mathbf s, \mathbf a^g) \ge \delta$). Using the classifier objective, the individual hinge loss for this specific catastrophic sample yields: 
\begin{equation}
\label{eq:52}
\ell_{\mathrm{hinge}}(\mathbf s, \mathbf a^g)
= \max\big(0, 1 - (-1)(\delta)\big)
= 1 + \delta
\end{equation}
The true hinge risk $\mathcal{R}_{\mathrm{hinge}}(\omega)$, which is defined as the expectation over the true data distribution, acts as an upper bound on the probability of these severe errors. Applying Markov's inequality ($\mathbb{P}(X \ge a) \le \frac{\mathbb{E}[X]}{a}$) bounds the probability of this specific catastrophic failure mode:
\begin{equation}
\label{eq:53}
\mathbb{P} \big(C_\omega(\mathbf s, \mathbf a^g) \ge \delta \mid y(\mathbf s, \mathbf a^g) = -1 \big)
\le \frac{\mathcal{R}_{\mathrm{hinge}}(\omega)}{1 + \delta}
\end{equation}
Let $\epsilon_{\mathrm{class}} = \mathcal{R}_{\mathrm{hinge}}(\omega)$ denote the converged validation loss of the centralized classifier. The probability that a gated action executed with a safety margin $\delta$ leads to an unsafe transition is bounded by:
\begin{equation}
\label{eq:54}
\mathbb{P}\big(C_\omega(\mathbf s, \mathbf a^g) \ge \delta \mid  \mathbf s' \in \mathcal{U} \big)
\le \frac{\epsilon_{\mathrm{class}}}{1 + \delta}
\end{equation}
Training the recourse policies to map rejected behavior proposals deep into the safe region maximizes the margin $\delta$ and drives down $\epsilon_{\mathrm{class}}$. The bounds demonstrate the practical utility of the dual-policy architecture where the MADWMPG framework systematically minimizes the probability of catastrophic network violations.

The MADWMPG framework is illustrated in Fig. \ref{fig:madwmpg} and Algorithm \ref{algo:dwmpg} summarizes the training procedure. 

\begin{algorithm}[!t]
\caption{MADWMPG Training}
\label{algo:dwmpg}
\begin{algorithmic}[1]

\State Initialize behavior policies $\mu_n$ and recourse policies $\rho_n$, $n \in \mathcal{N}$.
\State Initialize critic $Z_\psi$, classifier $C_\omega$, and target networks.
\State Initialize replay buffer $\mathcal D$.
\State Load fixed expert dataset $\mathcal D_{\mathrm{ex}}$.
\For{each training episode}
    \For{$t=0,\ldots,T-1$}
        \For{$e=1,\ldots,N_E$}
            \State Obtain stochastic state
            $\mathbf s_{e,t}$.
            \State Generate $\mathbf a_{e,t}^b$
            using $\boldsymbol\mu$ and exploration noise.
            \State Apply PCSOP predictor--corrector.
            \State Evaluate gate $I_{e,t}$
            using $C_\omega$.
            \If{$I_{e,t}=0$}
                \State Generate $\mathbf a_{e,t}^r$
                using $\boldsymbol\rho$.
                \State Apply PCSOP predictor--corrector.
            \EndIf
            \State Select $\mathbf a_{e,t}^g$
            according to $I_{e,t}$.
            \State Execute $\mathbf a_{e,t}^g$;
            observe $r_{e,t}$, $c_{e,t}$,
            and $\mathbf s_{e,t+1}$.
            \State Compute safety label $y_{e,t}$.
            \State Store transition $\xi_{e,t}$
            in $\mathcal D$.
        \EndFor

        \State Sample historical batch
        $\mathcal B_t$ from $\mathcal D$.
        \State Collect recent batch
        $\mathcal B_t^{\mathrm{rec}}$
        of $kN_E$ transitions.
        \State Form augmented batch
        $\widetilde{\mathcal B}_t
        =\mathcal B_t\Vert\mathcal B_t^{\mathrm{rec}}$.
        \State Sample expert batch
        $\mathcal B_t^{\mathrm{ex}}$
        from $\mathcal D_{\mathrm{ex}}$.

        \State Construct gated target actions
        $\bar{\mathbf a}_{j+1}^{g}$.
        \State Compute distributional targets
        $\mathcal Y_j^{(\ell)}$.
        \State Update $Z_\psi$ by minimizing
        $\mathcal L_Z$.
        \State Update $C_\omega$ by minimizing
        $\mathcal L_C$.
        \State Update $\mu_n$ by minimizing
        $\mathcal L_\mu$, $\forall n\in\mathcal N$.
        \State Update $\rho_n$ by minimizing
        $\mathcal L_\rho$, $\forall n\in\mathcal N$.
        \State Soft-update target policies
        and critic.
    \EndFor
\EndFor

\end{algorithmic}
\end{algorithm}

\section{Training and Simulation Results}
\begin{figure}[!t]
  \centering
  \includegraphics[width=1\linewidth]{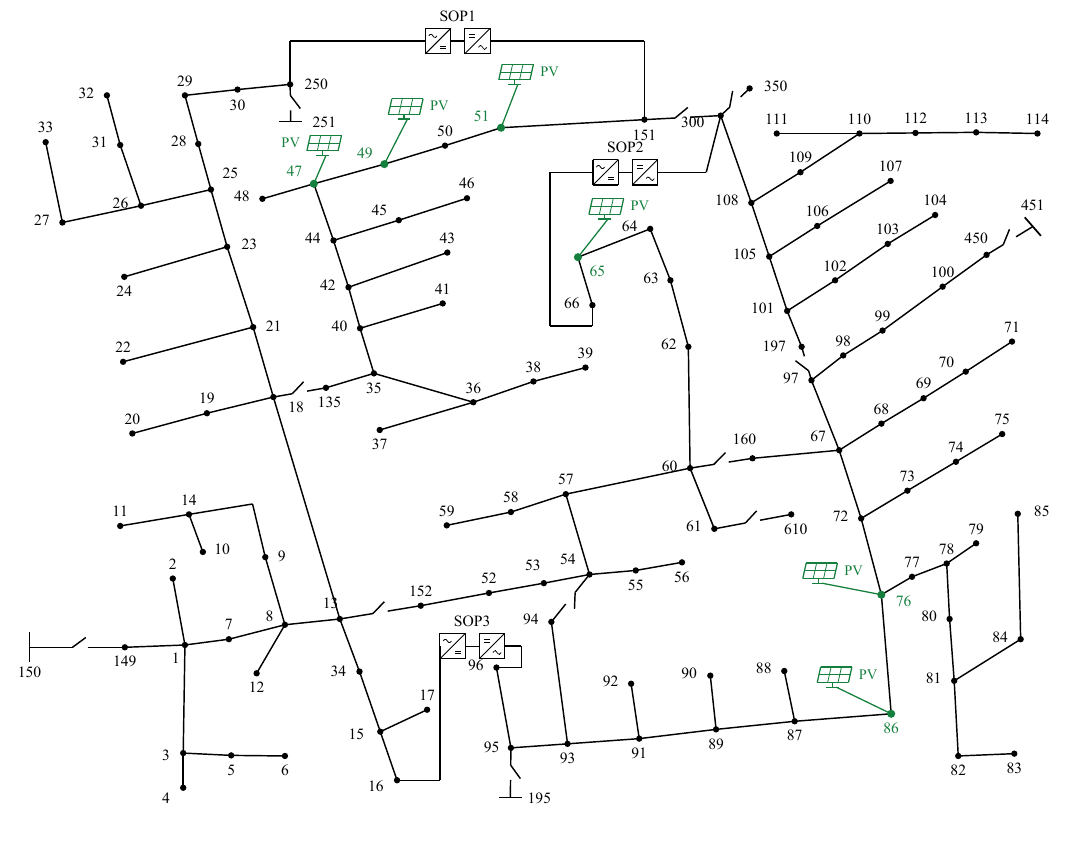}
  \caption{IEEE 123-bus test-feeder with DERs and PCSOPs.}
  \label{fig:ieee123bus}
\end{figure}

This section presents the results on evaluation of the MADWMPG framework. The framework is trained and tested on a modified IEEE 123-bus test feeder integrated with three spatially distributed PCSOPs. The set of dispatchable PCSOPs comprises two two-terminal three-phase PCSOPs and a two-terminal two-phase PCSOP. The IEEE 123-bus test feeder is a comprehensive low-voltage distribution system with characteristics that make optimization challenging. The topology comprises a mixed-phase configuration with three-, two-, and single-phase laterals branching from the main feeder. The radial network hosts several solar DERs with asymmetric phase loadings leading to unbalanced operation. To evaluate the performance of the framework on real-world operational patterns, the load data is adapted from the Norwegian 124-bus radial electric power distribution system dataset \cite{sperstadReferenceDataSet2023}, and the PV solar generation data is obtained from the UNISOLAR open dataset project \cite{wimalaratneUNISOLAROpenDataset2022}. The load data consists of a year of nodal demand time series, whereas the solar generation dataset provides two years of recorded time series, only one year of which is retained. Eight weeks (56 days) of uniformly-spaced operational data is extracted from each dataset as training data, with an additional eight days of data, spread over the year, isolated as a test set. The uniform spacing of the extracted training data captures seasonal patterns, while the test set enables extensive out-of-sample benchmarking of the framework. This configuration yields an approximate 87:13 train-test split. A visual representation of the system is depicted in Fig. \ref{fig:ieee123bus}. 

\begin{figure}[!t]
  \centering
  \includegraphics[width=1\linewidth]{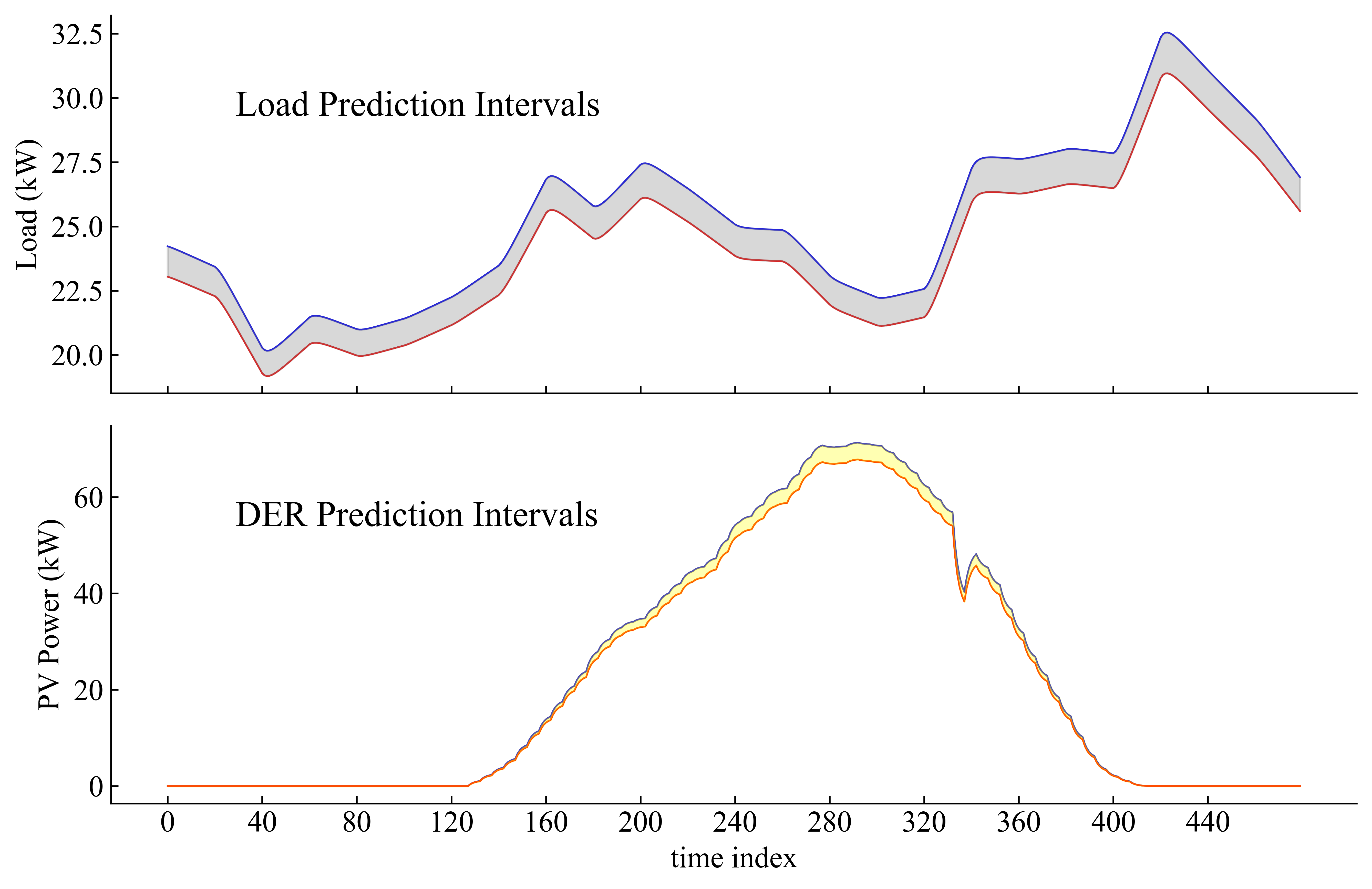}
  \caption{Bus load and DER prediction intervals over a day.}
  \label{fig:load_der_profile}
\end{figure}

Mapping the extracted data to the test feeder involves normalizing the operational profiles and multiplying them by the nominal bus load parameters defined by the IEEE network. The PV generation profiles are similarly adapted for the represented DERs. Although SOPs possess rapid-response capabilities, existing literature typically assumes slower dispatch intervals ranging from 15 to 60 minutes. To demonstrate high-resolution control while maintaining computational tractability, the proposed framework operates on a 3-minute (\(T\)=480) dispatch cycle. While MADWMPG can be readily extended to sub-minute dispatch rates, the associated training complexity scales nonlinearly with temporal resolution. The proposed framework is designed to address operational stochasticity using distributional critics. To simulate uncertainty in nodal demand and DER injections, the training environment incorporates time-varying, next-step prediction intervals generated by a standard probabilistic quantile forecasting model trained on the operational data. Fig. \ref{fig:load_der_profile} illustrates the prediction intervals of a typical bus load and PV power injections. The framework deploys parallelized power flow simulators to enhance training efficiency. 

The expert demonstrator first solves the ACOPF problem defined in (\ref{eq:12}) to generate a set of reference trajectories. Given the simulated uncertainty, the ACOPF solver optimizes only the nominal realization of the demand and DER profiles, neglecting the broader stochastic scenario tree. Consequently, it is statistically unlikely that the agent will traverse the exact expert-provided state trajectories during training. For a successful daily dispatch, solution times vary between 11 and 25 minutes. Furthermore, the solver frequently stalls or hits non-convergence timeouts, rendering this traditional approach unsuitable for quasi-real-time operation.

\begin{table}[!t]
\begin{center}
\caption{System Ratings.}
\label{tab:1}
\begin{tabular}{| c | c |}
\hline
Parameter & IEEE 123-bus \\
\hline
System voltage& 4.16 kV\\
\hline
Nominal system load& 3.49+j1.92 MVA\\
\hline
DER total peak generation& 0.75 MW\\
\hline
DER buses& 47, 49, 51, 65, 76, 86\\
\hline
PCSOP capacity per phase& 0.5 MVA\\
\hline
$\Gamma$& 0.02\\
\hline
PCSOP buses& (250,151), (300,66), (16,96)\\
\hline
PCSOP phases& (abc,bca), (bca,cab), (c,b)\\
\hline
Voltage limits (p.u.)& $\mathrm{V_{\min}} = 0.97$, $\mathrm{V_{\max}} = 1.03$\\
\hline
\end{tabular}
\end{center}
\end{table}

\begin{table}[!t]
\begin{center}
\caption{Training Hyperparameters.}
\label{tab:2}
\setlength{\tabcolsep}{3pt}
\begin{tabular}{| c | c | c | c | c | c |}
\hline
Parameter & MADWMPG & DDPG & TD3PG & RCPO & PID-$\lambda$ \\
\hline
Hidden layers& \multicolumn{5}{c|}{3}\\
\hline
Neurons per layer& \multicolumn{5}{c|}{512}\\ 
\hline
$|\mathcal B_t|$& \multicolumn{5}{c|}{512}\\
\hline
Noise ($\sigma$)& \multicolumn{5}{c|}{0.1}\\
\hline
$\gamma$& \multicolumn{5}{c|}{0.99}\\
\hline
Training steps& \multicolumn{5}{c|}{13440}\\
\hline
$N_E$& \multicolumn{5}{c|}{8}\\
\hline
$k_s$& \multicolumn{5}{c|}{0.375}\\
\hline
$k_v$& \multicolumn{5}{c|}{1.158}\\
\hline
$k_\theta$& \multicolumn{5}{c|}{0.497}\\
\hline
Optimizer& \multicolumn{5}{c|}{Adam}\\
\hline
Critic rate& 1e-3 & 1e-3 & 1e-3 & 1e-3 & 1e-3 \\
\hline
Policy rate& 2e-4 & 2e-4 & 2e-4 & 1e-4 & 1e-4 \\
\hline
$M$& 10 & - & - & - & - \\
\hline
$w_1^b$& 1.0 & - & - & - & - \\
\hline
$w_2^b$& 0.9 & - & - & - & - \\
\hline
$w_3^b$& 0.25 & - & - & - & - \\
\hline
$w_1^r$& 0.9 & - & - & - & - \\
\hline
$w_2^r$& 1.0 & - & - & - & - \\
\hline
$\lambda$ rate& - & - & - & 1e-4 & - \\
\hline
$K_p$& - & - & - & - & 5e-2 \\
\hline
$K_I$& - & - & - & - & 5e-3 \\
\hline
$K_D$& - & - & - & - & 5e-2 \\
\hline 
\end{tabular}
\end{center}
\end{table}

The proposed framework is implemented using the PyTorch deep learning library. The expert demonstrator and the power flow routines are implemented using the PYOMO optimization modeling language and solved using the IPOPT solver. The agent is trained on a personal computer with an Apple M4 Pro 14-core processor, 48 GB of system memory, and a 20-core GPU. Table \ref{tab:1} lists the system ratings that dictate the power flow dynamics. The voltage requirements in Table \ref{tab:1} imply that the framework's performance is benchmarked under tighter operating conditions.

\subsection{Performance Comparisons}
The performance of the MADWMPG framework is benchmarked against contemporary DRL and state-of-the-art SRL algorithms, in addition to the ACOPF baseline. Each of the baseline DRL agents is augmented with the proposed physics-based action generation module; without this augmentation, they are unable to navigate the complex state-action space of the problem, incurring excessive penalties that hinder learning convergence. The conventional DRL methods include DDPG and TD3PG, two of the most widely utilized deterministic policy algorithms. The SRL methods include the Reward Constrained Policy Optimization (RCPO), and the Proportional-Integral-Derivative Lagrangian (PID-$\lambda$) algorithm. RCPO is a primal-dual Lagrangian SRL method which forms the basis for many of the SRL contributions in power systems applications \cite{suReviewSafeReinforcement2025,chenPhysicsShieldedMultiAgentDeep2023,wangSafeOffPolicyDeep2020}. The PID-$\lambda$ is a state-of-the-art SRL method rooted in classical feedback control that generalizes the primal-dual Lagrangian SRL framework. To the best of the authors' knowledge, this work represents the first application of the PID-$\lambda$ framework to power distribution network control, serving to bridge the gap between general safe reinforcement learning literature and applied power systems optimization.            
A grid-search method is used to obtain the set of hyperparameters on the training set. The scaling factors $k_s$, $k_v$, and $k_\theta$ in (\ref{eq:5}) are selected such that they provide informative learning gradients at the lower end of the reward scale. The learning parameters are summarized in Table \ref{tab:2}, and the agents are trained using their respective optimal hyperparameters. Furthermore, to consistently benchmark the safety performance of the evaluated DRL methods, this paper introduces the Risk-Adjusted Constraint Violations (RACV) metric:
\begin{equation}
\label{eq:55}
\mathrm{RACV} = \left( \sum_{t=1}^{T} \frac{1}{N_E} \sum_{e=1}^{N_E} c_{e,t} \right) \left( \frac{1}{T} \sum_{t=1}^{T} \frac{1}{N_E} \sum_{e=1}^{N_E} k_{e,t} \right)
\end{equation}
where $k_{e,t}$ is the discrete number of voltage violations in environment $e$ at step $t$. The RACV tracks the cumulative average violation cost per environment $e$ and the cumulative average frequency of violations per environment over the operation horizon $T$. Traditional cumulative penalty metrics often obscure the distinction between sustained minor constraint breaches and rare catastrophic failures. The RACV metric explicitly decouples the cumulative magnitude of voltage violations from their empirical frequency over the dispatch horizon. By scaling the average daily violation cost by the empirical probability of violation occurrence, RACV penalizes policies that exhibit chronic insecurity, providing a representative measure of closed-loop operational reliability.  

\begin{figure}[!t]
\centering
\includegraphics[width=1\linewidth]{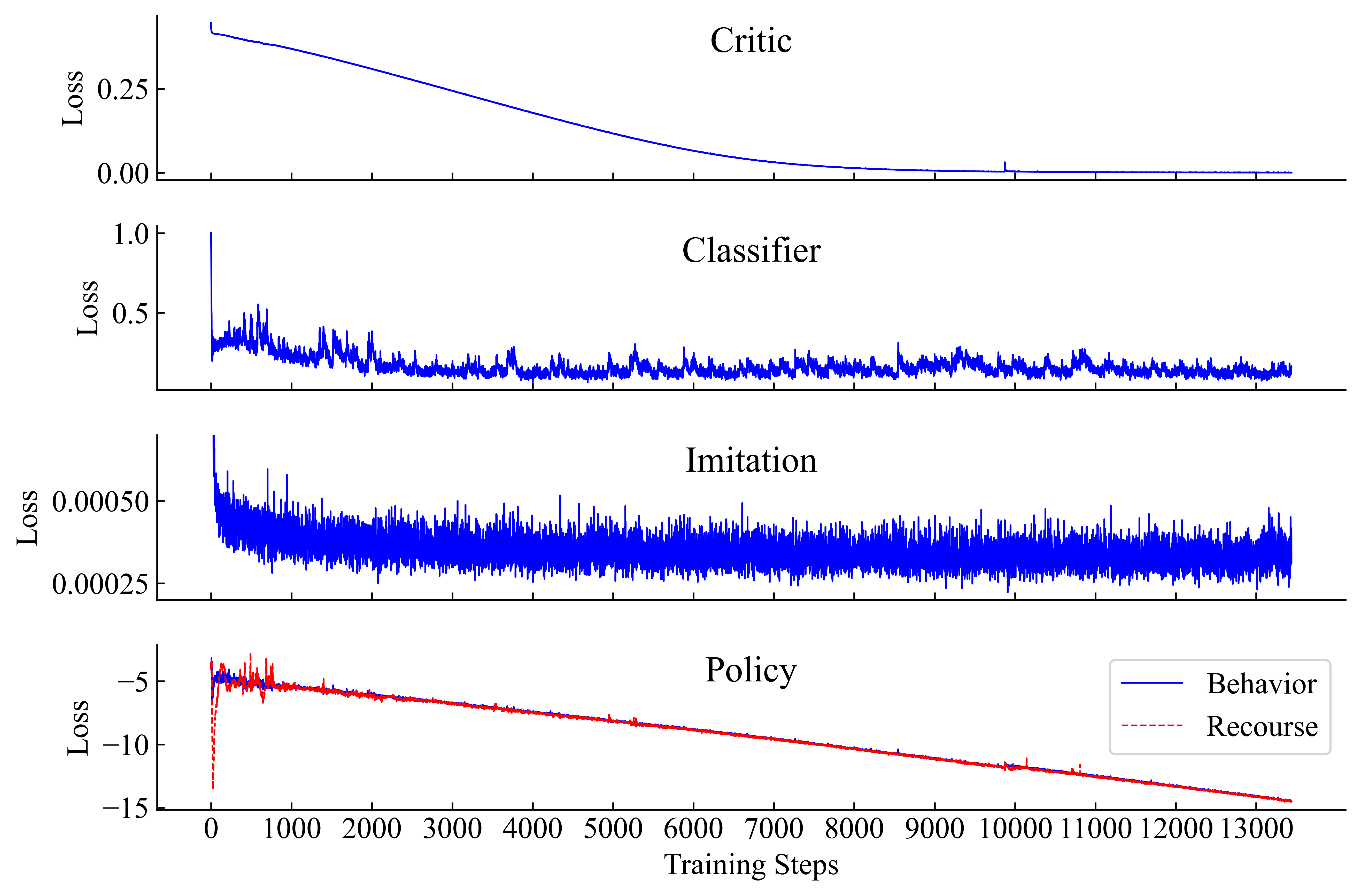} 
\caption{Critic, classifier, imitation, and policy loss components during MADWMPG training.}
\label{fig:training_losses}

\vspace{1.5em} 

\includegraphics[width=1\linewidth]{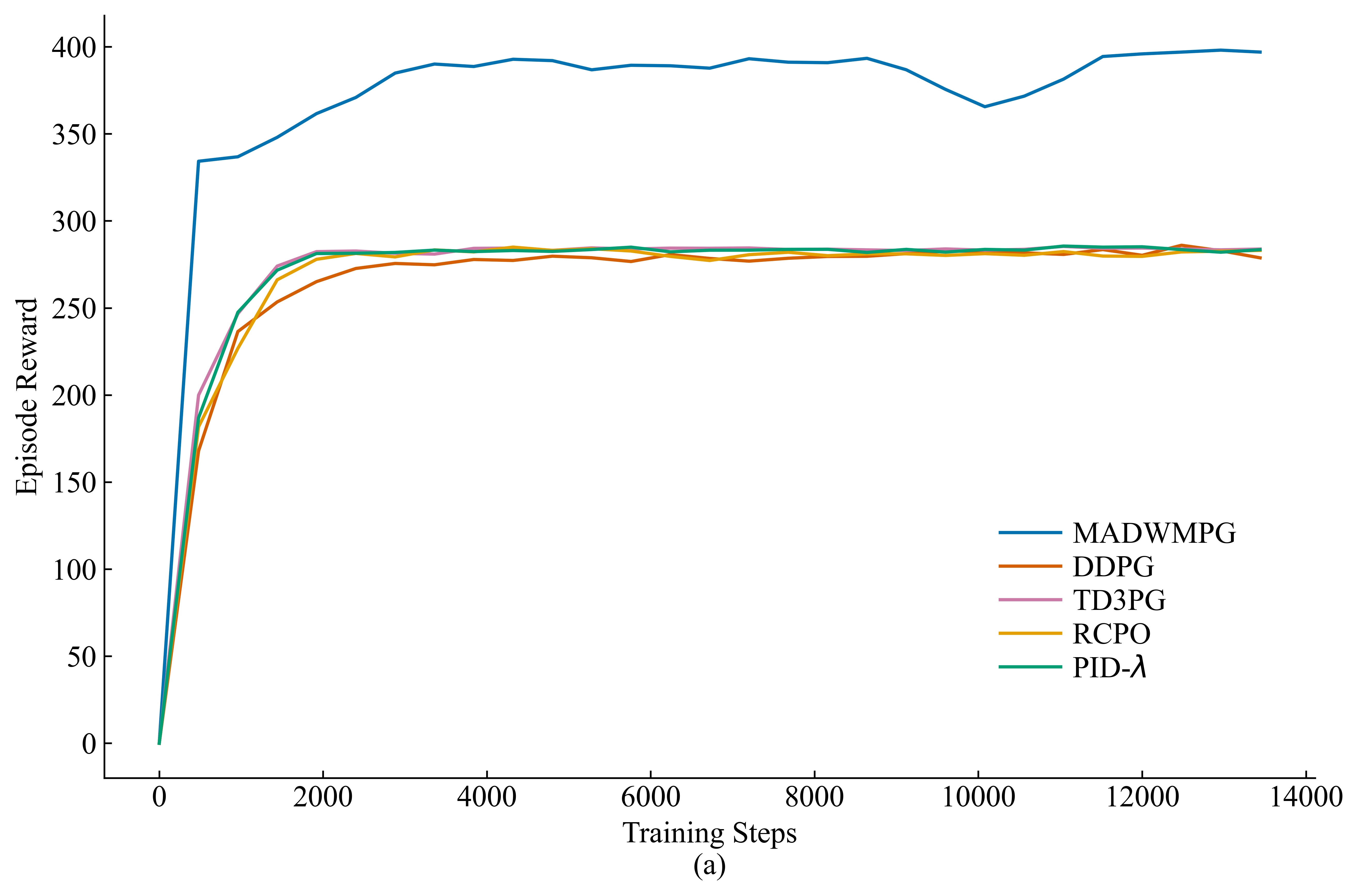} 
\includegraphics[width=1\linewidth]{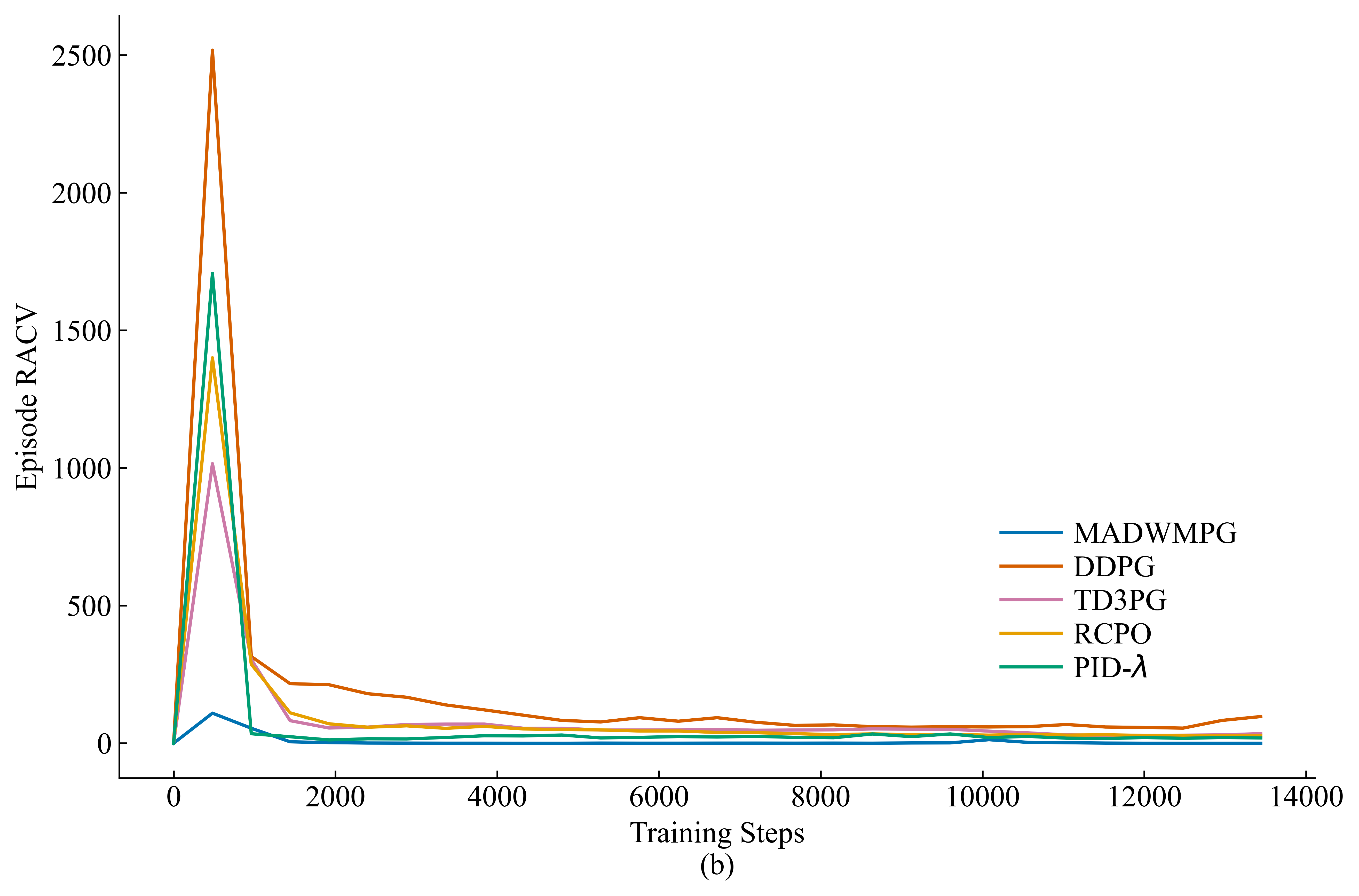} 
\caption{MADWMPG, DDPG, TD3PG, RCPO, and PID-$\lambda$ (a) rewards and (b) episodic RACV scores during training.}
\label{fig:rewards_and_racv}
\end{figure}

\begin{figure}[!t]
\centering
\includegraphics[width=1\linewidth]{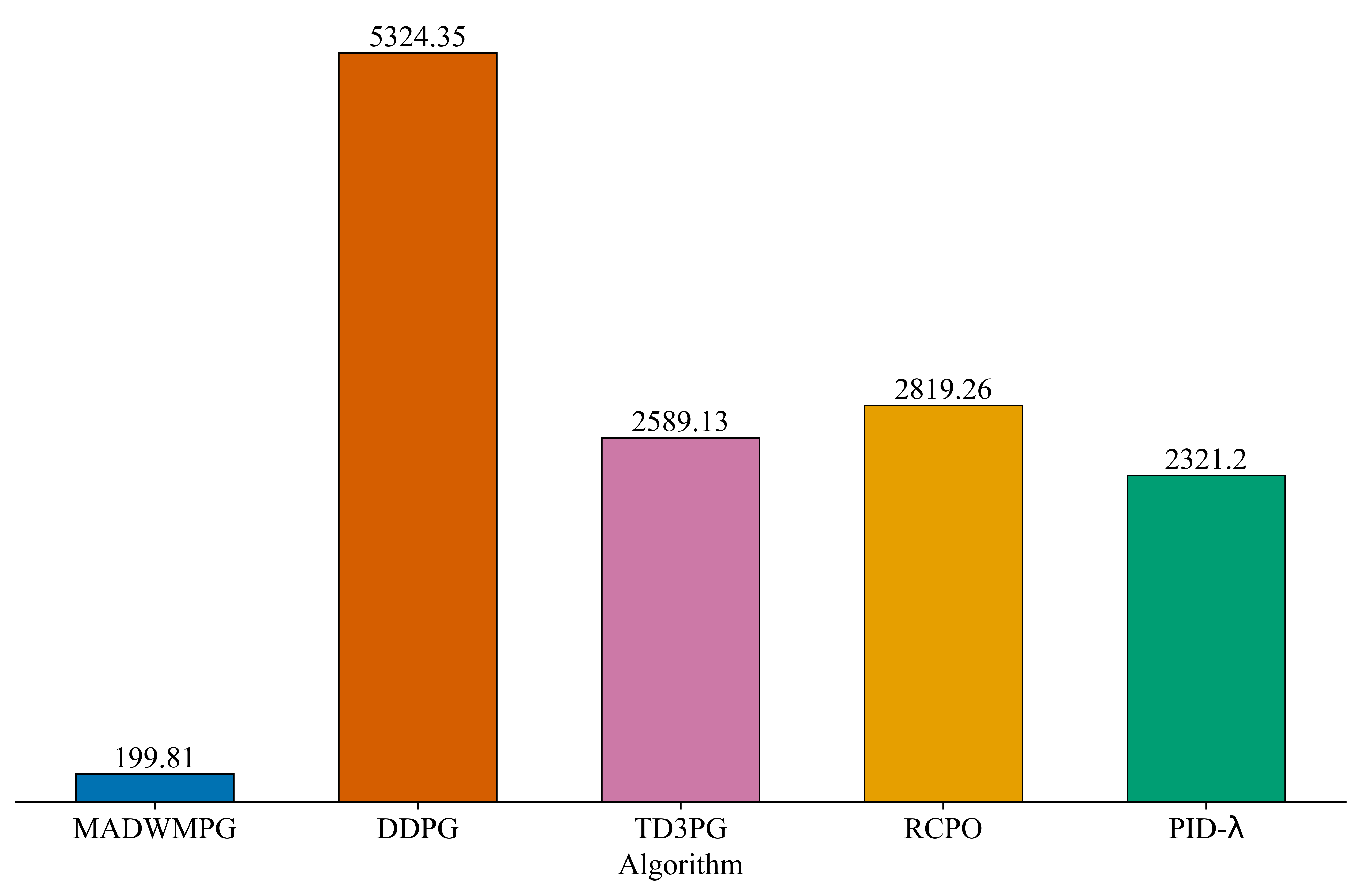} 
\caption{MADWMPG, DDPG, TD3PG, RCPO, and PID-$\lambda$ cumulative training RACV scores.}
\label{fig:training_cumulative_racv}
\end{figure}

Fig. \ref{fig:training_losses} shows the convergence of the distributional critic, wide-margin classifier, IL, and behavior and recourse policies' loss functions during MADWMPG training. The learning dynamics of the classifier and policy set are characteristic of the framework's closed-loop gating mechanism. The initial reduction in the classifier loss indicates that the agent is able to detect unsafe behavior proposals and pivot to the recourse actions. The recourse policies, however, have not learned a comprehensive safety mapping in the first 1000 training steps. The early oscillations in the recourse policy loss are indicative of this misalignment, which leads to new state visitations that the classifier is unfamiliar with. This accompanies a brief increasing trend observed in the classifier loss after its initial decline. The cycle repeats as the classifier loss converges over the training period. Beyond 2000 steps, the recourse actions harmonize with classifier interventions and the behavior policies that are trained in conjunction with expert-led imitation learning. The IL component of the policy loss converges quickly because of the static set of expert trajectories. The stochastic sampling of next-step states makes it statistically unlikely for the agent to experience grid states that are identical to the expert's, thereby reducing interference with critic-led training. In contrast to typical DRL critic training, the distributional critic loss converges smoothly over the training period as it tracks the full distribution of policy returns. Fig. \ref{fig:rewards_and_racv}(a) depicts the evolution of MADWMPG's episodic rewards during training compared to the baseline DRL methods. The four baseline methods exhibit typical exploratory convergence during training, which usually lags behind their deterministic inference scores. Among the baselines, the TD3PG and PID-$\lambda$ methods achieve the highest converged training scores. The proposed framework, however, significantly diverges from this trend, prominently outperforming the evaluated algorithms in converged training reward. The superior performance of MADWMPG compared to the roster of DRL methods can be attributed to the distributional critic, the closed-loop interactions of the WCBR module and explorations in the vicinity of the Expert's actions. The expert-provided reference rewards are not included here because the state trajectories of the agents are generally out-of-sample with respect to the expert dataset \(\mathcal D_{\mathrm{ex}}\).

The training rewards, however, mask a subtle detail that motivates the development of the MADWMPG framework, as depicted in figures \ref{fig:rewards_and_racv}(b) and \ref{fig:training_cumulative_racv}. Fig. \ref{fig:rewards_and_racv}(b) illustrates the dynamic exploratory safety performance of each method using the episodic RACV scores. Early in the training process, both the conventional and SRL methods incur frequent and large constraint violation costs, as reflected in the RACV trend. The PID-$\lambda$ method exhibits the sharpest improvement in exploratory safety within the first 2000 training updates. The conventional augmented DDPG method considerably underperforms in constraint compliance throughout the training. Interestingly, the safety dynamics of RCPO, a primal-dual SRL method, are observed to closely mirror those of TD3PG. This trend underscores the brittleness of integral-Lagrangian methods, which require extensive fine-tuning of the dual-variable learning rate. The MADWMPG framework clearly outperforms all other methods despite exploratory pressure. The wide-margin classifier and recourse policies effectively implement exploration-shaping that helps the agent circumvent frequent constraint violations early on. Fig. \ref{fig:training_cumulative_racv} summarizes the safety performance of all methods during training using the cumulative RACV scores. Over the course of the training cycle, MADWMPG yields an improvement of over 96\% and 91\% compared to augmented DDPG and the closest competing SRL alternative, respectively. It should be noted that the proposed method's superior safety performance stems inherently from the framework's recourse architecture rather than solely from expert guidance. The safety guarantee of behavior action proposals under expert imitation recedes when the actions are augmented with exploratory noise. Therefore, the agent relies on the robust classification of proposed actions and the recourse policies to safely explore the environment.

The proposed framework is tested extensively using an out-of-sample test set comprising 8 days, or 3840 dispatch instances. Table \ref{tab:3} summarizes the test scores of each DRL method compared to the Expert. The Expert, which represents the ideal performance baseline obtained by deterministically solving the ACOPF program, yields the highest average step reward score on the test set. This is closely followed by the proposed framework, which outperforms all other data-driven methods in terms of both application score and constraint satisfaction. The PID-$\lambda$ method achieves the best RACV score among the baseline DRL methods, but its score is still $\sim$13$\times$ higher than that of MADWMPG. To contextualize the safety performance of the proposed method, a cumulative RACV score of 0.006 corresponds to an average grid-wide per-instance constraint cost of $2.35 \times 10^{-3}$ p.u. and a mean per-bus-phase voltage deviation on the order of $\pm$$\mathrm{}{10^{-6}}$ p.u.    

\begin{figure}[!t]
\centering
\includegraphics[width=\linewidth]{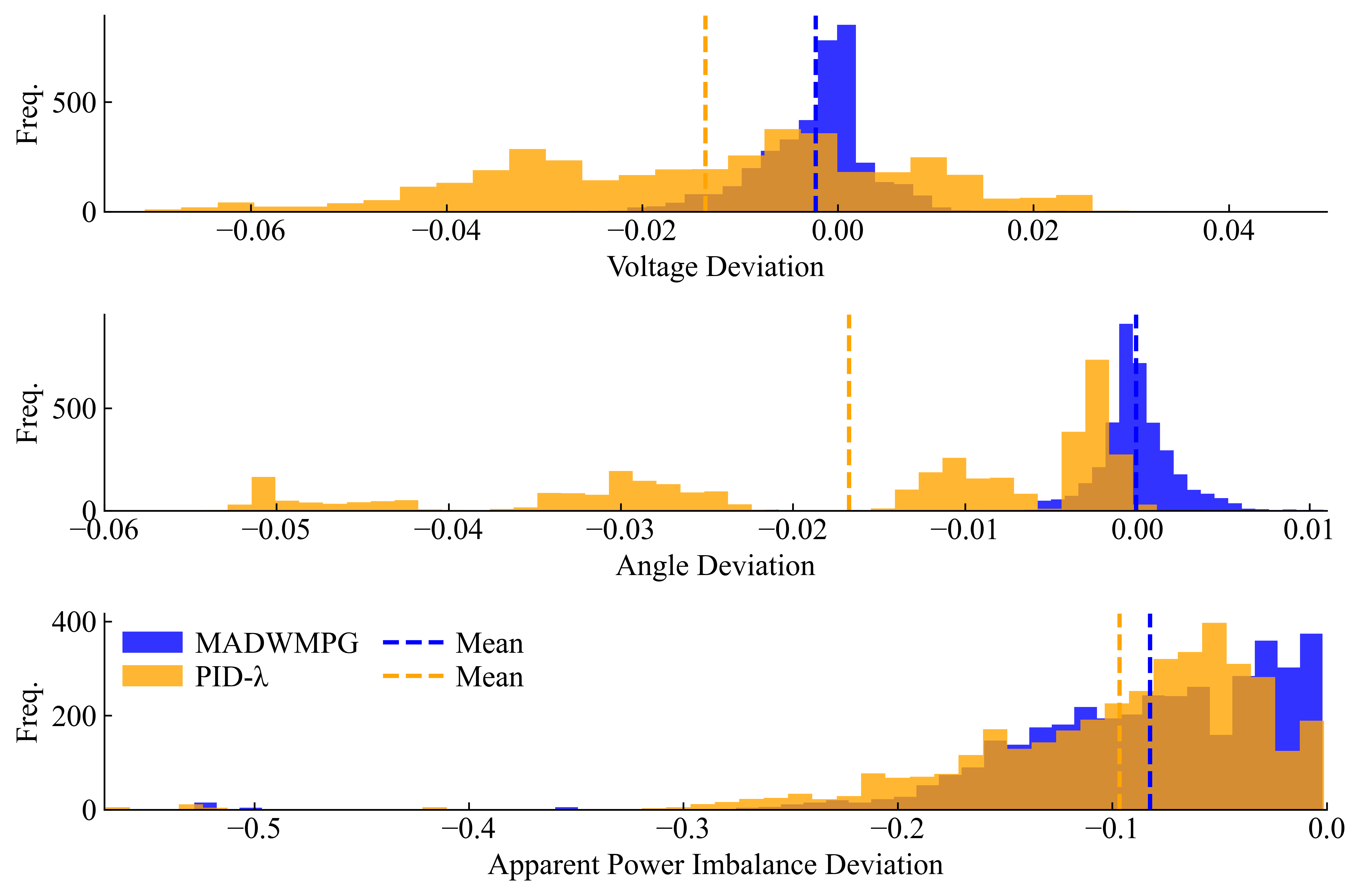}
\caption{Histograms of differences in reward components w.r.t. expert.}
\label{fig:test_component_histograms}
\end{figure}

\begin{figure}[!t]
\centering
\includegraphics[width=1\linewidth]{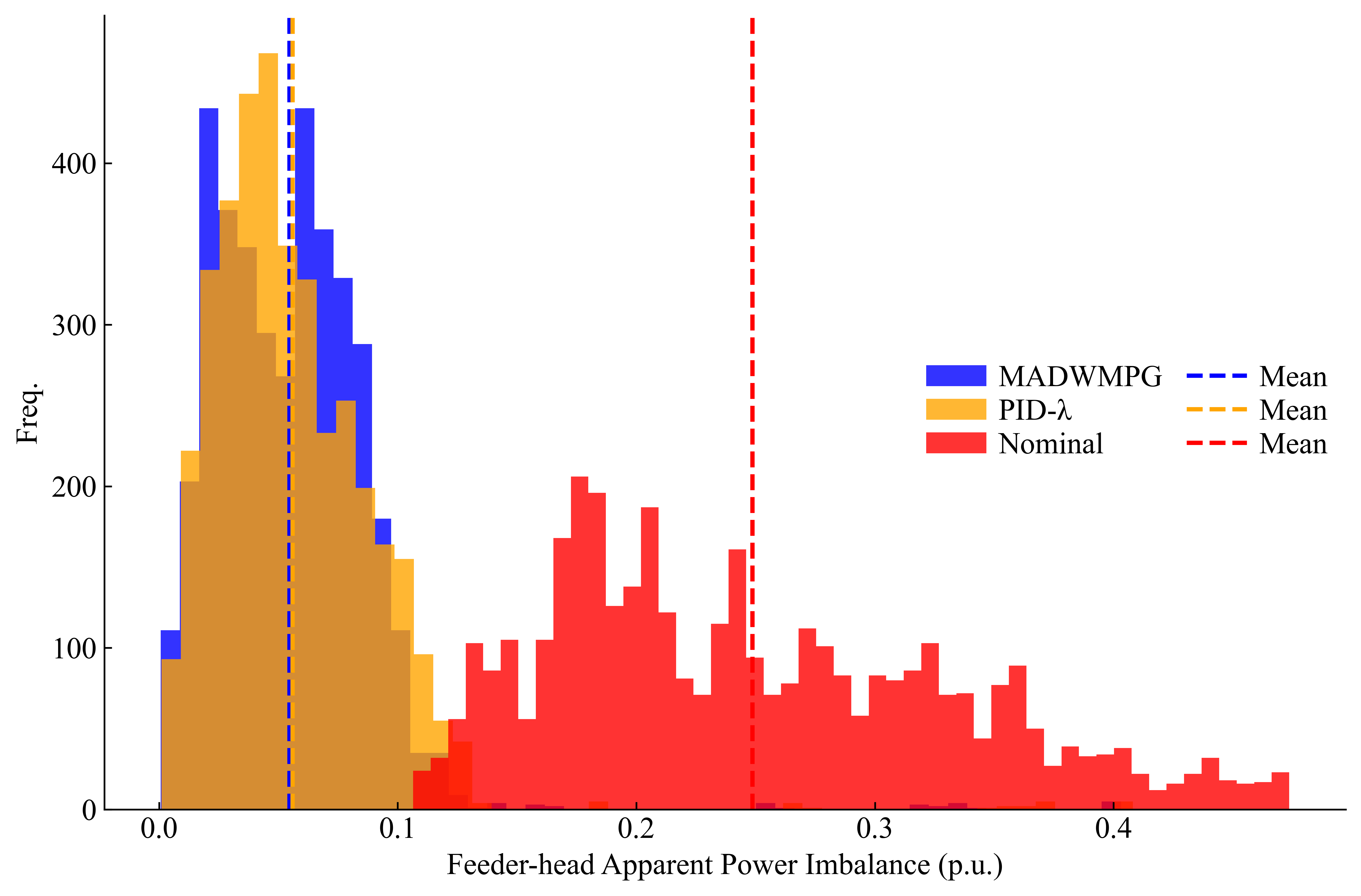} 
\caption{Feeder-head apparent power imbalance.}
\label{fig:feederhead_histogram}
\end{figure}

\begin{figure}[!t]
\centering
\includegraphics[width=1\linewidth]{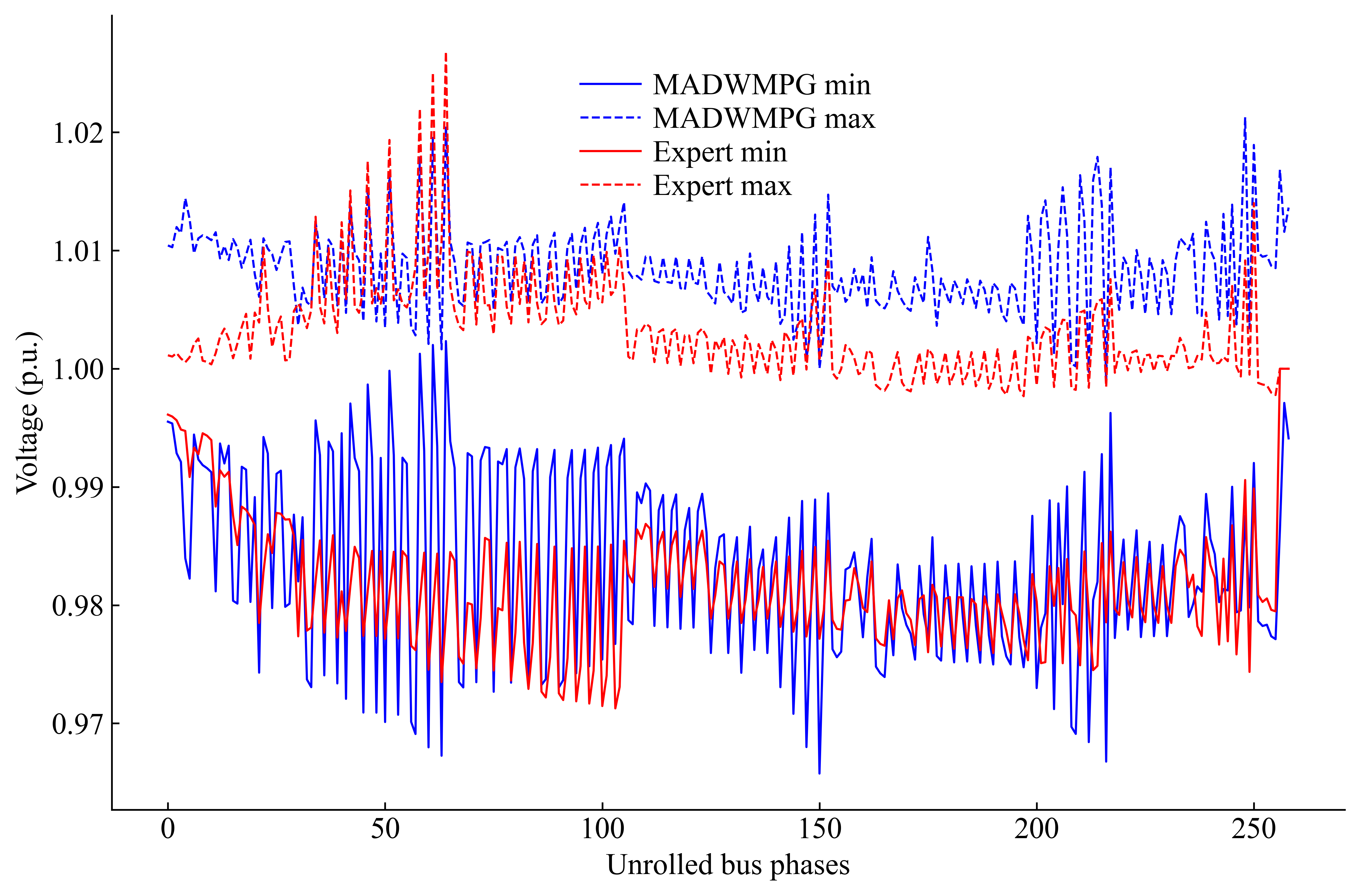} 
\caption{MADWMPG v. Expert voltage performance over the test days.}
\label{fig:test_voltages}
\end{figure}

\begin{table}[!t]
\begin{center}
\caption{Algorithm Performance Comparison.}
\label{tab:3}
\begin{tabular}{| c | c | c |}
\hline
Algorithm & Average Test Step Reward & Cumulative RACV \\
\hline
MADWMPG & 0.957 & 0.006 \\
\hline
DDPG & 0.929 & 35.010 \\
\hline
TD3PG & 0.935 & 4.571 \\
\hline
RCPO & 0.933 & 5.103 \\
\hline
PID-$\lambda$ & 0.933 & 0.076 \\
\hline
Expert & 0.989 & - \\
\hline
\end{tabular}
\end{center}
\end{table}

The performance of MADWMPG is further evaluated and compared to both the competing SRL method, PID-$\lambda$, and the Expert. To quantify the performance comparison more effectively, the histograms of the difference between MADWMPG and the Expert's individual reward components, i.e., voltage magnitude deviation, voltage angle distortion, and root-node apparent power imbalance, are shown in Fig. \ref{fig:test_component_histograms}. The histograms of the performance of the PID-$\lambda$ agent are also included along with the mean values of the differential distributions. Figures \ref{fig:test_component_histograms}(a) and \ref{fig:test_component_histograms}(b) illustrate the ability of the proposed agent in closely tracking the performance of the expert in both voltage regulation and voltage imbalance on the test set. The spread of the reward distribution toward positive differentials indicates that MADWMPG outperforms the expert in those instances. Infact, in the first two components the agent outperforms the expert in 37\% and 44\% of the test instances, respectively. The mean values of the framework's reward differentials remain close to the expert reference. The performance of the agent on feeder-head power imbalance mitigation presents an interesting dynamic and warrants a closer examination. All tested DRL methods report a wider spread in this reward component, which is indicative of the optimization complexity of data-driven grid imbalance mitigation applications. This is not a limitation of MADWMPG itself but exemplifies the design tradeoffs between state feature representations, the value of perfect information, and the curse of dimensionality. When solving for the optimal dispatch, the ACOPF explicitly leverages the grid admittance matrix and computes the node voltages, branch currents, and lateral power flows in order to optimize the aggregate global feeder-head imbalance. The RL agent, on the other hand, does not have access to the internal state variables and is forced to statistically approximate the underlying grid physics using the limited bus net-injection features. The dense state representation utilized in DRL lacks topological context and can be enhanced, in theory, by either augmenting the local features with branch flow variables or overhauling the policy design with advanced learning architectures. In practice, a severalfold increase in state dimensionality adversely affects agent performance due to the curse of dimensionality, whereas advanced learning methods such as attention Transformers are computationally expensive. Despite the relative spread in Fig. \ref{fig:test_component_histograms}(c), the agent successfully tracks the performance of the expert, yielding a mean reward component differential of 0.08 alongside infrequent instances that exhibit a larger drop in the component score. This is demonstrated in Fig. \ref{fig:feederhead_histogram}, which illustrates the distribution of the absolute difference in apparent power flow between each phase at the feeder head, compared to the baseline case where the PCSOPs are dispatched solely to operate the grid within the nominal voltage range. On the test set, the MADWMPG agent restricts the apparent power flow imbalance to within 0.1 p.u. in 96\% of the test instances at the root node. The PID-$\lambda$ method lags behind the proposed framework across all facets of the application. It is noteworthy that the idealized performance of the ACOPF-based expert is largely aspirational. The nonlinear, nonconvex ACOPF formulation involving PCSOPs cannot guarantee convergence or strict bounds on solution times in real-time applications. 

The performance difference between MADWMPG and the expert in terms of mitigating voltage imbalance is negligible. To better illustrate the relative performance difference over the 192 hours of test set operation, Fig. \ref{fig:test_voltages} shows the minimum and the maximum voltages at each phase of each bus. Note that the bus numbers in the IEEE 123-bus network have been relabeled for convenience in the unrolled bus-phase axis. As seen in this figure, the agent maintains the grid voltages within the predefined range similarly to the expert, and the difference between them is not considerable. The agent successfully maintains voltage security, with rare constraint violations accounting for 0.05\% of the test instances and a maximum reported phase voltage deviation of $\pm$0.004 p.u. This indicates the ability of the agent to adapt to operational stochasticity and verifies the hypothesis that the proposed architecture generalizes to unseen instances.

\subsection{Computation Time and Inference Modes}
Following training, the MADWMPG operates deterministically in a decentralized manner with superior computational speed at the inference stage. The trained critic and experience replay samples are traditionally discarded. An additional advantage of the proposed framework is that it also allows for centralized execution by retaining the wide-margin classifier and recourse policies for enhanced grid security. Training the agent requires 192 minutes on average using the parallelized environments and more than 800 minutes without parallelization. When operating in decentralized mode, inference requires an average time of 6 milliseconds, whereas centralized execution that involves safety assessment and recourse actions requires an average time of 11 milliseconds. This is in sharp contrast to the ACOPF solver, which, depending on the loading and DER profile, requires 11 to 25 minutes for each successful solution of the daily optimal dispatch for the IEEE 123-bus network. Furthermore, this inference time tends to increase as the system size or the temporal resolution increases. The system size or the dispatch rate is not a contributing factor to the inference time of the proposed framework, which is a key advantage of DRL models over traditional optimization techniques.

\section{Conclusion}
This paper proposes the Multi-Agent Distributional Wide-Margin Policy Gradient (MADWMPG) framework - a dual-policy hybrid Imitation Learning and Safe Reinforcement Learning (IL-SRL) method designed for coordinated dispatching of distributed PCSOPs to mitigate voltage and power flow imbalance in unbalanced three-phase distribution networks. The framework directly addresses the challenges of safe operational dispatching of SOPs that arise from grid stochasticity as well as the complexity of interconnected active-reactive power constraints. The framework formulates the first recourse-as-classification problem and features a Wide-Margin Constraint Barrier Recourse (WCBR) module, which consists of a robust centralized state transition classifier and a set of recourse policies. Furthermore, the MADWMPG integrates a novel adaptation of distributional critics that track the full distribution of policy returns enabling agent robustness to operational volatility. It also introduces a novel recency-biased experience replay enhancement that aids in rapid adaptation to constraint violations. The paper also proposes hybrid rank-weighted dual-policy loss functions that leverage distributional value learning, safety action gating, and imitation learning to enable safe exploration during training and convergence to safe inference policies. Ultimately, the MADWMPG framework closes the research gap on generalized, differentiable, and structurally embedded action projection mechanisms for SOPs by utilizing a physics-based, zero-shot predictor-corrector module that ensures seamless dispatch integrity. Moreover, the framework supports both inference modalities of decentralized execution and centralized control when higher reliability is desired. To support this architecture, the paper derives probabilistic bounds on the safety performance of the framework.

The performance of MADWMPG is demonstrated to be superior to conventional DRL and SRL methods and highly competitive against the ACOPF-based expert, outperforming it in many instances during out-of-sample testing. The framework outperforms the baseline DRL methods in terms of both constraint satisfaction and policy returns. There are three primary reasons for this superior performance: the ability of the physics-based action predictor to generate SOP-compliant injections without relying on active learning, which frees the agent to focus on application objectives; the safety classification-based policy losses that steer the policies away from exploratory violations; and exploration in the vicinity of a limited set of expert-provided optimal actions, which leads to higher rewards. The success of the framework in sample-efficient and constraint-compliant training provides a promising direction for safety-critical data-driven applications in active distribution networks. In this context, we encourage the research community to consider the following open research gaps:
\begin{itemize}
\item Formulating the coordinated real-time safe dispatch of SOPs alongside multi-timescale control devices remains an open problem.
\item Despite the framework's near-optimal performance in mitigating power flow imbalance at the feeder head, out-of-sample testing reveals infrequent deviations from the optimal response. The root power flow in an ADN is a rigid global state variable that is dependent on all upstream line flows along the routing path. We surmise that augmenting the policy architecture with physics-informed topological learning is a more promising future direction than relying on denser state feature representations.
\end{itemize}
The framework training and test source codes will be available on GitHub.

\bibliographystyle{IEEEtran}
\bibliography{ref.bib}

\vfill

\end{document}